\documentclass[prl,twocolumn,lengthcheck]{revtex4-1}

\usepackage{amsmath}
\usepackage{amssymb}
\usepackage{graphicx}
\usepackage[usenames,dvipsnames]{color}
\usepackage{hyperref}
\usepackage{braket}

\newcommand{\operator}[1]{\ensuremath{\hat{#1}}}
\newcommand{\ic}{\ensuremath{\mathrm{i}}}

\newcommand{\rbra}[1]{\ensuremath{(#1|}}
\newcommand{\rket}[1]{\ensuremath{|#1)}}
\newcommand{\rbraket}[1]{\ensuremath{(#1)}}

\begin{document}

\title{Time-dependent variational principle for quantum lattices}

\author{Jutho Haegeman$^1$}
\author{J. Ignacio Cirac$^{2}$}
\author{Tobias J. Osborne$^3$}
\author{Iztok Pi\v{z}orn$^4$}
\author{Henri Verschelde$^{1}$}
\author{Frank Verstraete$^{4,5}$}
\affiliation{$^1$Ghent University, Department of Physics and Astronomy, Krijgslaan 281-S9, B-9000 Ghent, Belgium\\
$^2$Max-Planck-Institut f\"ur Quantenoptik, Hans-Kopfermann-Str. 1, Garching, D-85748, Germany\\
$^3$Leibniz Universit\"at Hannover, Institute of Theoretical Physics, Appelstrasse 2, D-30167 Hannover, Germany\\
$^4$University of Vienna, Faculty of Physics, Boltzmanngasse 5, A-1090 Wien, Austria\\
$^5$C.N. Yang Institute for Theoretical Physics, SUNY, Stony Brook, NY 11794-3840, USA}

\begin{abstract}
	We develop a new algorithm based on the \emph{time-dependent variational principle} applied to \emph{matrix product states} to efficiently simulate the real- and imaginary time dynamics for infinite one-dimensional quantum lattice systems. This procedure: (1) is argued to be optimal; (2) does not rely on the Trotter decomposition and thus has no Trotter error; (3) explicitly preserves all symmetries and conservation laws; and (4) has low computational complexity. The algorithm is illustrated using both an imaginary time and a real-time example.
\end{abstract}

\maketitle

The density-matrix renormalization group (DMRG) is arguably the most powerful tool available for the study of one-dimensional strongly interacting quantum lattice systems \cite{dmrg}. The DMRG --- now understood as an application of the variational principle to \emph{matrix product states} (MPS) \cite{mps} --- was originally conceived as a method to calculate ground-state properties. However, there has been a recent explosion of activity, spurred by insights from quantum information theory, in developing   powerful extensions allowing the study of, \textit{e.g.}, finite-temperature properties, higher-dimensional systems, and nonequilibrium physics via real-time evolution \cite{tdmrg}. The simulation of nonequilibrium properties with the DMRG was first attempted in \cite{tdmrg2}, but modern implementations are based on the \emph{time-evolving block decimation} algorithm (TEBD) and relatives \cite{tebd}. 

At the core of a TEBD algorithm lies the \emph{Lie-Trotter decomposition} for the propagator $\exp(\ic dt \operator{H})$, which splits it into a product of \emph{local unitaries}. This product can then be dealt with in a parallelised and efficient way: when applied to an MPS one obtains another MPS with larger bond dimension. To proceed one then \emph{truncates} the MPS description by discarding irrelevant variational parameters. This is such a flexible idea that it has allowed even the study of the dynamics of infinite translation-invariant lattice systems via the \emph{iTEBD} \cite{itebd}. Despite its success the TEBD has some  drawbacks: (1) the truncation step may not be optimal; (2) conservation laws, e.g.\ energy conservation, may be broken; and (3) symmetries, e.g., translation invariance, are broken (although translation invariance by two-site shifts is retained for nearest neighbor Hamiltonians). The problem is that when the Lie-Trotter step is applied to the state --- stored as an MPS --- it leaves the variational manifold and a representative from the manifold must be found that best approximates the new time-evolved state. There are a variety of ways to do this based on diverse distance measures for quantum states but implementations become awkward when symmetries and conservation laws are brought into account.  

In this Letter we introduce a new algorithm to solve the aforementioned problems --- intrinsic to the the TEBD --- without an appreciable increase in computational cost. The resulting imaginary time algorithm quickly converges towards the globally best uniform MPS (uMPS) approximation for translational invariant ground states of strongly correlated lattice Hamiltonians, and the corresponding real-time evolution evolves an initial state without violating energy conservation for constant Hamiltonians, or the conservation of any other quantities dictated by symmetry. The complexity of our approach can be made to scale as $D^3$, comparable with current implementations, where $D$ is the bond dimension of the uMPS. 

We now introduce the variational manifold $\mathcal{M}_{\text{uMPS}}$ of uniform MPS for an infinite lattice of spin-$d/2$ degrees of freedom, parameterized via
\begin{equation}
\textstyle
\ket{\psi(A)}=\sum_{\{s_{k}\}=1}^{d} v_{\mathrm{L}}^{\dagger}\big(\prod_{n\in\mathbb{Z}} A^{s_{n}} \big)v_{\mathrm{R}}\ket{\mathbf{s}} ,\label{eq:umpsdef}
\end{equation}
where $\ket{\mathbf{s}} \equiv \ket{\ldots s_{1}s_{2}\ldots}$ and $v_{\mathrm{L}}$ and $v_{\mathrm{R}}$ are two $D$-dimensional vectors, which are presently argued to be irrelevant. The variational parameters $A$ comprise the set of $D\times D$ matrices $A^{s}$ ($s=1,2,\ldots,d$) and are denoted via a $dD^{2}$ vector with entries $A^{i}=A^{s}_{\alpha,\beta}$, with $i=(\alpha,s,\beta)$ a collective index. The uMPS variational manifold has a \emph{gauge invariance}: replacing $A^s \mapsto G A^s G^{-1}$ for invertible $G$ results in an identical state. We do not fix the gauge and simply assume that $A^{s}$ are completely general complex matrices. We do, however, assume that the \emph{transfer matrix} $E= \sum_{s=1}^d A^{s}\otimes \bar{A}^{s}$ has precisely one eigenvalue $1$ with corresponding left and right eigenvectors $\rbra{l}$ and $\rket{r}$ of length $D^2$, to which we can associate $D\times D$ matrices $l$ and $r$, respectively, by simply reshaping them. These matrices are Hermitian and positive and assumed to have full rank. We choose the normalization so that $\rbraket{l|r}=\mathrm{Tr}(l r)=1$. In addition, we assume that all other eigenvalues of $E$ lie strictly within the unit circle, i.e.\ the spectral radius of $E-\rket{r}\rbra{l}$ is smaller than $1$. These conditions allow one to write for any local operator $\operator{O}$ acting on $n$ contiguous sites:
\begin{displaymath}
\begin{split}
&O(\overline{A},A)=\braket{\psi(\overline{A})|\operator{O}| \psi(A)}/\braket{\psi(\overline{A})|\psi(A)}=\\
&\ \ \rbraket{l| \sum_{s,t=1}^{d} O_{t_{1}\ldots t_{n},s_{1}\ldots s_{n}} (A^{s_{1}}\cdots A^{s_{n}}) \otimes (\overline{A}^{t_{1}}\cdots \overline{A}^{t_{n}})|r}.
\end{split}
\end{displaymath}
The boundary vectors $v_{\mathrm{L}}$ and $v_{\mathrm{R}}$ do not feature in normalized expectation values and thus do not contain any variational degrees of freedom.

\begin{figure}
\includegraphics[width=0.75\columnwidth]{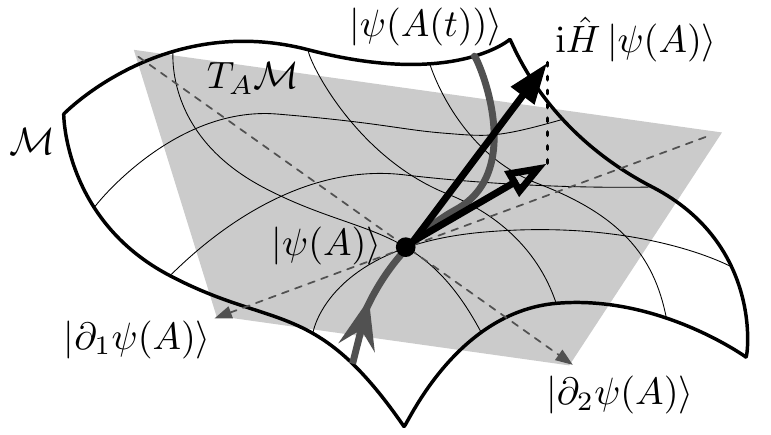}
\caption{An illustration of our construction: the wireframe surface represents the variational manifold $\mathcal{M}=\mathcal{M}_{\text{uMPS}}$ embedded in state space, with the black dot a point representing a uMPS $\ket{\psi(A)}$. The rotated gray square represents the tangent plane $T_{A}\mathcal{M}$ to $\mathcal{M}$ in $\ket{\psi(A)}$, with two generally non-orthogonal coordinate axes $\ket{\partial_{1}\psi(A)}$ and $\ket{\partial_{2}(A)}$ displayed as dotted lines. The arrow with solid head is the direction $\ic\operator{H}\ket{\psi(A)}$ of time evolution, and the arrow with open head represents the vector that best approximates $\ic\operator{H}\ket{\psi(A)}$ within the tangent plane. The gray curve is the optimal path $\ket{\psi(A(t))}$ which follows the vector field generated by these vectors with open head throughout $\mathcal{M}$.}\label{fig:manifold}
\end{figure}

Denote a translation invariant nearest-neighbour Hamiltonian as $\operator{H}=\sum_{n\in\mathbb{Z}} \operator{T}^{n} \operator{h}^{n}\operator{T}^{-n}$, where $\operator{T}$ is the shift operator and $\operator{h}$ acts non-trivially only on sites zero and one. We now try to approximate the time evolution generated by $\operator{H}$ of a uMPS $\ket{\psi(A)}$ without ever leaving the variational manifold of uMPS with fixed bond dimension $D$, by introducing a time-dependent parameterisation $A(t)$. Insertion into the time-dependent Schr\"odinger equation results in $\dot{A}^{i} \ket{\partial_{i} \psi(A(t))}=-\ic \operator{H}\ket{\psi(A(t))}$, where we denote $\partial_{i}$ for $\partial/\partial A^{i}$. Whereas the left hand side (LHS) is a linear combination of the tangent vectors $\ket{\partial_{i} \psi(A(t))}$ that span the tangent plane $T_{A}\mathcal{M}_{\text{uMPS}}$, the right hand side (RHS) is a general vector in Hilbert space and this equation does not have an exact solution for $\dot{A}^{i}$. The best approximation is obtained by minimizing
\begin{displaymath}
\lVert \dot{A}^{i} \ket{\partial_{i} \psi(A(t))}+\ic \operator{H}\ket{\psi(A(t))}\rVert.
\end{displaymath}
The minimum is found by orthogonally projecting the evolution vector $\operator{H}\ket{\psi(A(t))}$ onto the tangent plane, as illustrated in Fig.~\ref{fig:manifold}. The resulting solution is determined by
\begin{equation}
\braket{\partial_{\overline{\jmath}}\psi|\partial_{i} \psi}\dot{A}^{i} =-\ic \braket{\partial_{\overline{\jmath}}\psi|\operator{H}|\psi}, \label{eq:timeevomps}
\end{equation}
where the argument $A(t)$ in every vector has been omitted for the sake of brevity. The LHS of Eq.~\eqref{eq:timeevomps} contains the $dD^{2}\times dD^{2}$ Gram matrix of the tangent vectors $G_{\overline{\imath},j}(\overline{A},A)=\braket{\partial_{\overline{\imath}}\psi(\overline{A})|\partial_{j} \psi(A)}$.
Expressions for this Gram matrix and the vector in the RHS of Eq.~\eqref{eq:timeevomps} are best derived using the explicit form for the tangent vector 
$B^{i}\ket{\partial_{i} \psi(A)}= \sum_{n\in\mathbb{Z}}\operator{T}^{n} \sum_{\{s_{k}\}=1}^{d} v_{\mathrm{L}}^{\dagger}\left(\cdots A^{s_{-1}} B^{s_{0}} A^{s_{1}}\cdots \right)v_{\mathrm{R}}\ket{\mathbf{s}}$, and are given by
\begin{multline*}
\overline{B'}^{\overline{\imath}} G_{\overline{\imath},j}B^{j}=
|\mathbb{Z}|\Big[\rbraket{l|E^{B}_{B'}|r}\\+\rbraket{l|E^{A}_{B'}(1-E)^{-1}E^{B}_{A}|r}+ \rbraket{l|E^{B}_{A}(1-E)^{-1}E^{A}_{B'}|r}\\
+(|\mathbb{Z}-1)|\rbraket{l|E^{A}_{B'}|r}\rbraket{l|E^{B}_{A}|r}\Big],\\
B^{\overline{\imath}}\braket{\partial_{\overline{\imath}}\psi|\operator{H}|\psi}=|\mathbb{Z}|\Big[\rbraket{l|H^{AA}_{AB}|r}+\rbraket{l|H^{AA}_{BA}|r}\\ +\rbraket{l|H^{AA}_{AA}(1-E)^{-1}E^{A}_{B}|r} +\rbraket{l|E^{A}_{B}(1-E)^{-1}H^{AA}_{AA}|r}\\+(|\mathbb{Z}|-2) \rbraket{l|E^{A}_{B}|r}\rbraket{l|H^{AA}_{AA}|r}\Big],
\end{multline*}
where $E^{A}_{B}=\sum_{s=1}^d A^{s}\otimes \overline{B}^{s}$ (note the identity $E=E^{A}_{A}$) and $H^{AB}_{CD}=\sum_{s,t,u,v=1}^d \braket{s,t|\operator{h}|u,v}(A^{u}B^{v}) \otimes(\overline{C}^{s}\overline{D}^{t})$. In these expressions, $(1-E)^{-1}$ should be interpreted as the pseudo-inverse of $(1-E)$, i.e.\ it produces zero when acting on the left or right eigenvector of $E$ with eigenvalue $1$: $\rbra{l}(1-E)^{-1}=0=(1-E)^{-1}\rket{r}$. The overall factors $|\mathbb{Z}|$ are a consequence of the infinite volume of our system and cancel, as they appear both in the LHS and RHS of Eq.~\eqref{eq:timeevomps}. The additional divergent terms on the last line of the brackets would disappear if we restricted ourselves to tangent vectors that are orthogonal to the uMPS itself, such that $\braket{\psi(A)|\partial_{i} \psi(A) } B^{i}= |\mathbb{Z}| \rbraket{l|E^{B}_{A}|r}=0$. Indeed, the tangent plane contains the state itself, since $A^{i}\ket{\partial_{i} \psi(A)}=|\mathbb{Z}|\ket{\psi(A)}$. However, a change in that direction would change the norm or phase of the state, which is not a desired effect. 

This construction can also be derived from an action principle and is known as the \emph{time-dependent variational principle} (TDVP) \cite{tdvp,supplementarymaterial}. The resulting TDVP equations [Eq.~\eqref{eq:timeevomps}] can be shown to be sympletic \cite{tdvpsymplectic}. Hence they respect energy conservation as well as conservation of all constants of motion, such as the expectation value of generators of symmetries. Since only expectation values occur in the equations of motion, one can use techniques familiar from DMRG, including the decomposition of the matrices $A^{i}$ into irreducible representations of the relevant symmetry group. Further, this approach is manifestly translation invariant. For time-reversal invariant operators the TDVP equations are also invariant under time reversal (see \cite{trottertimereversal} for a Trotter-based approach that recovers time reversal invariance). This approach does not require any truncation and is thus globally optimal within the manifold $\mathcal{M}_{\text{uMPS}}$.

Constructing the relevant quantities and solving Eq.~\eqref{eq:timeevomps} for $\dot{A}^{i}$ involve operations with a computational complexity of $\mathcal{O}(D^{6})$. Using an iterative method to implement $(1-E)^{-1}$ and then solving for $\dot{A}^{i}$ can reduce this to $\mathcal{O}(D^{3})$. However, the matrix $G_{\overline{\imath},j}$ is not invertible: because of the gauge invariance in the (u)MPS parameterisation, not all $dD^{2}$ tangent vectors are linearly independent. Defining the action of a $1$-parameter group of gauge transformations $G(\varepsilon)=\exp(\varepsilon X)$ as $A^{s}(\varepsilon)=G(\varepsilon) A^{s} G(\varepsilon)^{-1}$, we obtain that $dA^{s}/d \varepsilon =X A^{s} - A^{s}X$. Because of gauge invariance, there is no corresponding change in $\ket{\psi(A(\epsilon))}$ and thus $d \ket{\psi(A(\varepsilon))} /d\varepsilon = (d A^{i}/d \varepsilon) \ket{\partial_{i} \psi}=0$. Indeed, any vector $B_{X}^{i}$ defined by $B_{X}^{s}=X A^{s}- A^{s}X$ produces a zero norm state, evident when introducing it into the explicit form of $B^{i}\ket{\partial_{i}\psi(A)}$. The vectors $B^{i}_{X}$ thus span the null space of $G_{\overline{\jmath},i}$. Any vector $B$ in the tangent plane is gauge equivalent to $B'=B+B_{X}$, $\forall X\in \mathbb{C}^{D\times D}$. There are $D^{2}-1$ linearly independent choices of $B_{X}$, as we can easily prove by noting that $B_{X}=0$ requires that $\sum_{s=1}^{d} (A^{s})^{\dagger} l B_{X}^{s}=0=\sum_{s=1}^{d} (A^{s})^{\dagger} l X A^{s}-l X$. Since $E$ has a single eigenvalue $1$, and $l$ has full rank, the only solution to this equation is $X=1$. In order to invert $G_{\overline{\imath},j}$, we fix the gauge which eliminates $D^{2}-1$ components of $B$. Norm preservation (\textit{i.e.}\ $\rbraket{l|E^{B}_{A}|r}$) fixes one more component, resulting in a $(d-1)D^{2}$ dimensional tangent plane. 

While there are a variety of ways to fix the gauge of vectors in the tangent plane, different choices result in different effective Gram matrices with different condition numbers. By using the gauge fixing condition $\rbra{l} E^{B}_{A}=0$ --- which also includes norm preservation and imposes the condition that the eigenvalue and left eigenvector of the transfer matrix do not change to first order --- the effective Gram matrix reduces to $\overline{B'}^{\overline{\imath}}G_{\overline{\imath},j}B^{j}=|\mathbb{Z}|\rbraket{l|E^{B}_{B'}|r}$ and all non-local contributions are thus effectively canceled. Let us now explain how to exploit this result even further. We start by defining the $D\times dD$ matrix $L_{\alpha,(s\beta)}=[(A^{s})^{\dagger}l^{1/2}]_{\alpha\beta}$. Clearly, the null space of this matrix is $D(d-1)$ dimensional. Let the $Dd\times D(d-1)$ matrix $V_{L}$ with entries $[V_{L}]_{(\alpha s),\gamma}$ be a matrix of orthonormal basis vectors for this null space, which can be obtained from, \textit{e.g.}\ the singular value decomposition of $L$, and thus satisfies $L V_{L}=0$ and $V_{L}^{\dagger}V_{L}=1$. We also introduce the notation $V_{L}^{s}$ for the $D\times D(d-1)$ matrix with components $[V^{s}_{L}]_{\alpha,\gamma}=[V_{L}]_{(\alpha s), \gamma}$. If we now group the $(d-1) D^{2}$ independent components of $B$ in a $D(d-1)\times D$ matrix $x$, we can use a parameterisation $B(x)$ given by $B^{s}(x) = l^{-1/2} V_{L}^{s} x r^{-1/2}$. One can check that this parameterisation satisfies the left gauge fixing constraint $\rbra{l}E^{B(x)}_{A}=0$ since $V_{L}$ contains only null vectors of $L$, and that $\overline{B}^{\overline{\imath}}(x) G_{\overline{\imath}j}B^{j}(y) = |\mathbb{Z}| \mathrm{tr}[x^{\dagger} y]$, since the vectors in $V_{L}$ are orthonormal. Up to the overall diverging factor $|\mathbb{Z}|$ that cancels in the LHS and RHS of Eq.~\eqref{eq:timeevomps}, we have found a linear parameterisation $B(x)$ for which the effective Gram matrix is the unit matrix. This same parameterisation cancels the last two terms in $\braket{\partial_{\overline{\imath}} \psi|\operator{H}|\psi}$. The third term is still non-local, and requires the inversion of $1-E$. However, this is a pseudo-inverse as $E$ has a single eigenvalue $1$ and $1-E$ is thus singular. Let $\rbra{K}=\rbra{l}H^{AA}_{AA}(1-E)^{-1}$. We can safely replace $\rbra{l}H^{AA}_{AA}$ by $\rbra{l}H^{AA}_{AA}-h\rbra{l}$, where $h=\rbraket{l|H^{AA}_{AA}|r}$, since $\rbra{l}(1-E)^{-1}=0$. Then, by replacing $1-E$ with the non-singular matrix $1-E+\rket{r}\rbra{l}$, we iteratively solve for the $D\times D$ matrix $K$ from
\begin{displaymath}
K-\sum_{s=1}^{d} (A^{s})^{\dagger} K A^{s} + \mathrm{tr}[K r] l =  \left[\rbra{l} H^{AA}_{AA}\right] - h l
\end{displaymath}
with $\left[\rbra{l} H^{AA}_{AA}\right]  = \sum_{stuv} \braket{s t|\operator{h}| uv} (A^{s}A^{t})^{\dagger} l (A^{u}A^{v})$. Tracing this equation shows that $\mathrm{tr}[K r]=\rbraket{K|r}=0$ as required. Finally, we define the $D(d-1) \times D$ tensor $F$
\begin{multline*}
F=\sum_{s,t=1}^{d}(V^{s}_{L})^{\dagger} l^{1/2} C^{st} r (A^{t})^{\dagger} r^{-1/2}\\ +\sum_{s=1}^{d} (V^{s}_{L})^{\dagger} l^{-1/2} \left(\sum_{t=1}^{d} (A^{t})^{\dagger}l C^{ts}+K A^{s}\right) r^{1/2},
\end{multline*}
where $C^{st}= \sum_{uv} \braket{s t|\operator{h}| uv} A^{u}A^{v}$. This definiton allows to write $\lVert B^{i}(x)\ket{\partial_{i}\psi} - \operator{H}\ket{\psi}\rVert^{2}=|\mathbb{Z}| \mathrm{tr}\left[ x^{\dagger} x -x^{\dagger} F - F^{\dagger} x + \text{constant}\right]$.
This expression is minimized by choosing $x=x^{\ast}\stackrel{\Delta}{=}F$ and thus $\dot{A}^{i}=-\ic B(x^{\ast})$. Note that, thanks to the iterative solver, all steps can be performed in $\mathcal{O}(D^{3})$ computation time.

Having now an explicit construction of $\dot{A}^{i}$, the simulation of time evolution with the TDVP now boils down to integrating a set of non-linear coupled differential equations. The simplest numerical integrator is built on the Euler method and proceeds as follows. 
\begin{enumerate}
\item Construct $x^{\ast}=F$ from the previous paragraph.
\item Set $A(t+dt)=A(t) - \ic dt B(x^{\ast})$.
\item Fix the gauge and norm of $A$ by rescaling $A$.
\item Calculate the energy and evaluate the step, change the time step $dt$ if necessary.
\end{enumerate}
Step 3 is required since the gauge-fixing condition only fixes the norm and left eigenvector up to first order and higher order corrections are generally present. This simple implementation is already useful for finding ground states through imaginary time evolution ($dt \to -\ic d\tau$). The TDVP then produces the best approximation to a gradient descent in the full Hilbert space which should be contrasted to a pure gradient-descent in parameter space such as in \cite{gradientdescent}. For real-time evolution, a simple first-order Euler integrator does not  inherit the symplectic properties of the differential equations and a more advanced integrator (see \cite{supplementarymaterial}) should be used.

We now illustrate the power of our approach. Using imaginary time evolution with the simple Euler implementation of the TDVP we've obtained a uMPS approximation for the ground state of the $S=1$ Heisenberg antiferromagnet. The TDVP stops when $\braket{\partial_{\overline{\imath}}\psi|\operator{H}|\psi}=0$, which indeed signals a minimum in the energy expectation value. Since the gradient has zero length at the minimum, it automatically decreases in size as we approach it, and there is typically no need to reduce the size of the time step. This should be compared with the (i)TEBD case, where reduction of the time step, and thus automatic slowing down, is necessary to overcome the Trotter error. An ordinary laptop or pc allows one to find the ground state up to $D=1024$ in less than one hour (without exploiting symmetries), resulting in a ground state energy density $e=-1.4014840389712(2)$ obtained with step size $dt=0.1$. Since we can easily calculate the norm of the gradient as $\eta=\lVert x^{\ast}\rVert$, we can continue the evolution until $\eta$ has converged below a specified tolerance level. The convergence of the energy can be shown to be $\mathcal{O}(\eta^{2})$ and can already be far beyond machine precision. This allows a much more accurate localization of the energy minimum than with the ordinary variational principle based on convergence of the energy, and is useful to \textit{e.g.}\ obtain a very accurate convergence in the entanglement spectrum. The entanglement spectrum can offer valuable information but is not converged very accurately by other approaches (see \cite{haldane} for an example). Table~\ref{tab:schmidtheisenberg} shows how the first Schmidt values of the uMPS ground state for the Heisenberg chain at $D=128$, which was converged up to $\eta=10^{-10}$, accurately reproduce the degeneracy according to half-integral spin representations. Note that we can also asses the error of being confined to the manifold at any point in the evolution and derive from this a construction to optimally increase the bond dimension. Rather than starting from a random state at $D=1024$, we can progressively build better approximations at larger $D$. Details are given in \cite{supplementarymaterial}.

\begin{table}[t]
\begin{tabular}{|c|c|c|}
\hline
\textcolor{red}{0.6961989782} & \textcolor{blue}{0.0057700505} & \textcolor{PineGreen}{0.0014877669}\\
\textcolor{red}{0.6961989782} & \textcolor{blue}{0.0057700505} & \textcolor{PineGreen}{0.0014877669}\\
\cline{1-1}
\textcolor{blue}{0.0860988815} & \textcolor{blue}{0.0057700505} & \textcolor{PineGreen}{0.0014877669}\\
\textcolor{blue}{0.0860988815} & \textcolor{blue}{0.0057700505} & \textcolor{PineGreen}{0.0014877669}\\
\cline{2-2}
\textcolor{blue}{0.0860988815} & \textcolor{blue}{0.0016659093} & \textcolor{PineGreen}{0.0014877669}\\
\textcolor{blue}{0.0860988815} & \textcolor{blue}{0.0016659093} & \textcolor{PineGreen}{0.0014877669}\\
\cline{1-1}\cline{3-3}
\textcolor{red}{0.0200132616} & \textcolor{blue}{0.0016659093} & \textcolor{red}{0.0011065273}\\
\textcolor{red}{0.0200132616} & \textcolor{blue}{0.0016659093} & \textcolor{red}{0.0011065273}\\
\hline
\end{tabular}
\begin{tabular}{c}
Color labels:\\
\textcolor{red}{S=1/2}\\
\textcolor{blue}{S=3/2}\\
\textcolor{PineGreen}{S=5/2}\\
\end{tabular}
\caption{First 24 Schmidt values of the $D=128$ uMPS approximation for the ground state of the $S=1$ Heisenberg antiferromagnet. The degeneracy in the Schmidt spectrum as a result of $\mathsf{SU}(2)$ symmetry manifests itself, not by exploiting the symmetry, but rather by converging up to `state tolerance' $\eta=10^{-10}$.}
\label{tab:schmidtheisenberg}
\end{table}

\begin{figure}[b]
\includegraphics[width=0.9\columnwidth]{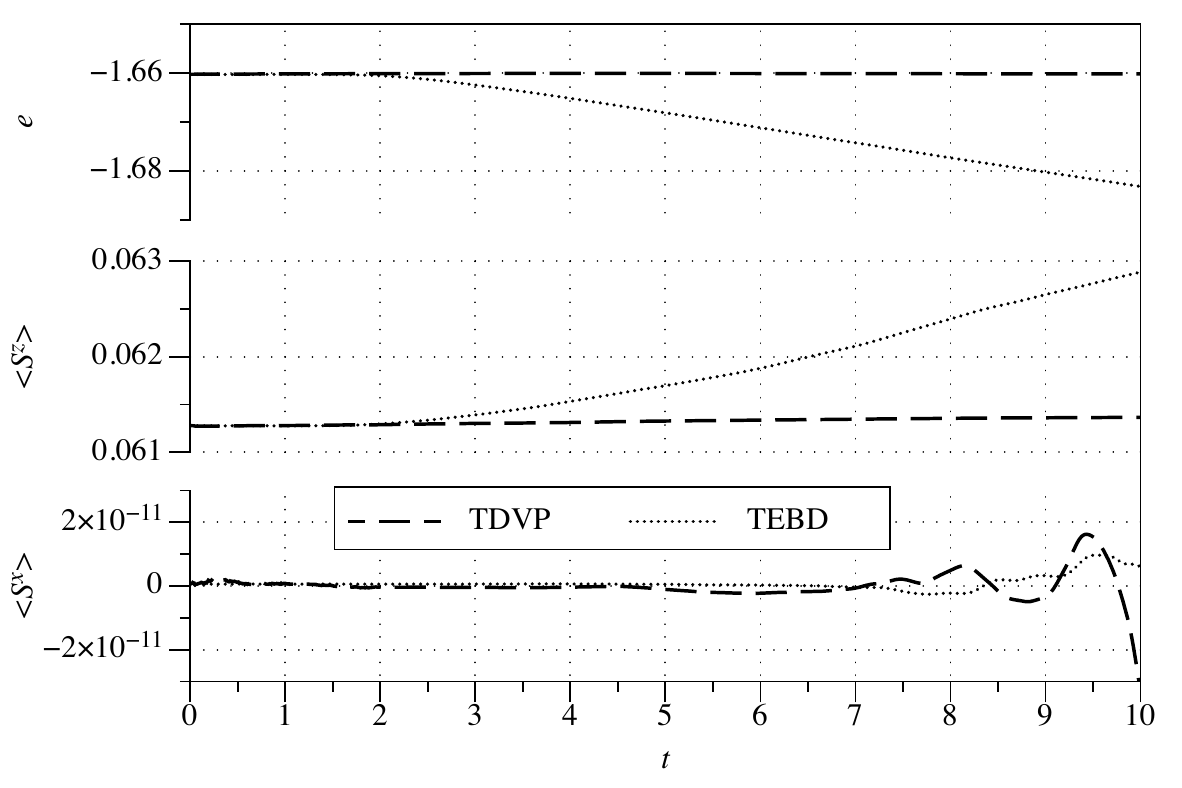}
\caption{Comparison of real-time simulation results at $D=128$ with time step $dt=5\times 10^{-3}$ for conserved quantities $e$ (energy density), $\braket{\operator{S}^{x}}$ and $\braket{\operator{S}^{z}}$ with TDVP (dashed lines) and TEBD (dotted lines).}
\label{fig:realtimeresults}
\end{figure}

Using the time-reversal invariant numerical integrator discussed in \cite{supplementarymaterial}, we can simulate a real-time evolution using the TDVP equations. We start with the $D=128$ uMPS ground state approximation of the $XX$-model with magnetic field $\mu=1/2$ along the $z$-axis, which is a critical model with non-zero magnetization $\braket{\operator{S}^{z}}\neq 0$, whereas $\braket{\operator{S}^{x}}=\braket{\operator{S}^{y}}=0$ due to the $\mathsf{U}(1)$ symmetry.  We evolve this state according to the critical $S=1/2$ Heisenberg antiferromagnet, so the expectation values $\braket{\operator{S}^{x,y,z}}$ should be conserved due to the $\mathsf{SU}(2)$ symmetry. Comparative results for the TDVP implementation and a second order, translation-invariant TEBD implementation based on \cite{tebdmpo} are shown in Fig.~\ref{fig:realtimeresults} and illustrate that TDVP is much more capable of describing the evolution of conserved quantities.

In this Letter we have introduced a new algorithm for simulating real and imaginary time evolution with (uniform) matrix product states. The algorithm is shown to be globally optimal within the variational manifold, while conserving all symmetries in the system.

\begin{acknowledgements}
Work suported by Research Foundation Flanders (JH), SFB projects, FoQuS and ViCoM, EU projects Quevadis, ERC grant QUERG and DFG-FG635. TJO acknowledges the support of the EU project COQUIT (Hannover).
\end{acknowledgements}


\begin{thebibliography}{99}
\bibitem{dmrg} 
S.R.~White, {Phys. Rev. Lett.} {\bf 69} 2863 (1992);
U.~Schollw{\"o}ck, {Rev. Mod. Phys.} {\bf 77} 259 (2005).

\bibitem{mps} M.~Fannes, B.~Nachtergaele, R.~F.~Werner, {Commun. Math. Phys.} {\bf 144}, 443 (1992); F.~Verstraete, J.~I.~Cirac, V.~Murg, {Adv. Phys.} {\bf  57}, 143 (2008); J.~I.~Cirac, F.~Verstraete, {J. Phys. A: Math. Theor.} {\bf  42},  504004  (2009); U.~Schollw{\"o}ck, {Annals of Physics} {\bf 326}, 96 (2011).

\bibitem{tdmrg}
U.~Schollw{\"o}ck and S.R.~White, in G.G.~Batrouni and D.~Poilblanc (eds.): {\it Effective models for low-dimensional strongly correlated systems}, AIP, New York (2006)

\bibitem{tdmrg2}
M.A.~Cazalilla and J.B.~Marst on, {Phys. Rev. Lett.} {\bf 88}, 
256403 (2002); {Phys. Rev. Lett.} {\bf 91}, 049702 (2003); H.G.~Luo, T.~Xiang and X.Q.~Wang, {Phys. Rev. Lett.} {\bf 91}, 049701 (2003). 

\bibitem{tebd}
G.~Vidal, {Phys. Rev. Lett.} {\bf 93}, 040502 (2004); S.R.~White and A.~Feiguin, {Phys. Rev. Lett.} {\bf 93}, 076401 (2004); A.J.~Daley, C.~Kollath, U.~Schollw\"ock and G.~Vidal, {J. Stat. Mech.: Theor. Exp.} (2004) P04005.

\bibitem{itebd}
G.~Vidal, {Phys. Rev. Lett.} {\bf 98}, 070201 (2007).

\bibitem{tdvp}
P.A.M.~Dirac, {Proc. Camb. Phil. Soc.} {\bf 26}, 376 (1930); P.W.~Langhoff, S.T.~Epstein and M.~Karplus, {Rev. Mod. Phys.} {\bf 44}, 602 (1972).

\bibitem{supplementarymaterial}
See supplementary material.

\bibitem{tdvpsymplectic}
A.K.~Kerman and S.E.~ Koonin, {Annals of Physics} {\bf 100}, 332 (1976); P.~Kramer and M.~Saraceno, {\it Geometry of the Time-Dependent Variational Principle in Quantum Mechanics}  (Springer-Verlag, Berlin) (1981).

\bibitem{trottertimereversal}
K. Ueda, C. Jin, N. Shibata, Y Hieida, T Nishino, arXiv:cond-mat/0612480 (2006).

\bibitem{gradientdescent}
B.~Pirvu, F.~Verstraete, and G.~Vidal, {Phys. Rev. B} {\bf 83}, 125104 (2011).

\bibitem{haldane}
J.~Zhao, D.N.~Sheng, F.D.M.~Haldane, arXiv:1103.0772 (2011).

\bibitem{tebdmpo}
B.~Pirvu, V.~Murg, J.I.~Cirac and F.~Verstraete, {New J. Phys.} {\bf 12}, 025012 (2010).

\end{thebibliography}
\end{document}


\title{Supplementary Material for\\
``Time-dependent variational principle for quantum lattices''}

\author{Jutho Haegeman$^1$}
\author{J. Ignacio Cirac$^{2}$}
\author{Tobias J. Osborne$^3$}
\author{Iztok Pi\v{z}orn$^4$}
\author{Henri Verschelde$^{1}$}
\author{Frank Verstraete$^{4,5}$}
\affiliation{$^1$Ghent University, Department of Physics and Astronomy, Krijgslaan 281-S9, B-9000 Ghent, Belgium\\
$^2$Max-Planck-Institut f\"ur Quantenoptik, Hans-Kopfermann-Str. 1, Garching, D-85748, Germany\\
$^3$Leibniz Universit\"at Hannover, Institute of Theoretical Physics, Appelstrasse 2, D-30167 Hannover, Germany\\
$^4$University of Vienna, Faculty of Physics, Boltzmanngasse 5, A-1090 Wien, Austria\\
$^5$C.N. Yang Institute for Theoretical Physics, SUNY, Stony Brook, NY 11794-3840, USA}

\maketitle
\onecolumngrid

This supplementary material discusses the following additional topics:
\begin{itemize}
\item A review of the time-dependent variational principle (TDVP) and its properties by deriving it from an action principle.
\item Additional details about the implementation, including a description of convergence and error measures and a complete description of the time-reversal invariant numerical integrator. 
\item A construction to dynamically increase the bond dimension of the uniform MPS.
\item A generalization of the proposed strategy to the case of non-uniform MPS on finite lattices.
\end{itemize}

\section{Review of the time-dependent variational principle}
In this section we review the general framework of the time-dependent variational principle, as can be found in [P.~Kramer and M.~Saraceno, {\it Geometry of the Time-Dependent Variational Principle in Quantum Mechanics}  (Springer-Verlag, Berlin) (1981)]. Recall that the time-dependent Schr\"odinger equation (TDSE) can be derived from extremizing an action functional $S\{\overline{\psi}(t),\psi(t)\}=\int_{t_{1}}^{t_{2}}\d t\, L(\overline{\psi}(t),\psi(t),t)$ with Lagrangian
\begin{equation}
L(\overline{\psi}(t),\psi(t),t)=\frac{\ic}{2}\braket{\psi(t)|\dot{\psi}(t)}-\frac{\ic}{2}\braket{\dot{\psi}(t)|\psi(t)}-\braket{\psi(t)|\operator{H}(t)|\psi(t)}.\label{eq:lagrangian}
\end{equation}
For the sake of brevity, we  henceforth omit the time dependence of $\operator{H}$ and thus of $L$. Stationarity of the action under independent variations of $\ket{\psi}$ and $\bra{\psi}$ in the full Hilbert space $\mathcal{H}$  yields the TDSE and its complex conjugate. But when we only have access to a subspace or manifold $\mathcal{M}\subset \mathcal{H}$, we can still use the calculus of variations with respect to this action to define a time evolution of states $\ket{\psi}\in \mathcal{M}$; this is the essence of the TDVP. Hereto, we restrict to variations in the tangent plane of $\mathcal{M}$ at the point $\ket{\psi}$. Assume that the manifold $\mathcal{M}$ can be parametrized as
\begin{equation}
\mathcal{M}=\{\ket{\psi(\vz)},\vz\in \mathbb{C}^{n}\},\label{eq:manifold}
\end{equation}
where we assume the dependence on the $n$ complex parameters $z^{i}$ to be analytic and explicitly denote the anti-analytic dependence $\bra{\psi(\ovz)}$. Furthermore, we introduce the notation $\partial_{i}$ for $\partial/\partial z^{i}$ and $\partial_{\overline{\jmath}}=\partial/\partial \oz^{\overline{\jmath}}$, where we  always use barred indices for the complex conjugate variables $\ovz$. Requiring stationarity of $S\{\ovz(t),\vz(t)\}$ with respect to a variation $\ovz(t)\to \ovz(t)+\overline{\vec{\delta z}}(t)$ results in the following Euler-Lagrange equations
\begin{equation}
\ic G_{\overline{\imath}j}(\ovz(t),\vz(t)) \dot{z}^{j}(t) = \braket{\partial_{\overline{\imath}} \psi(\ovz(t))|\operator{H}|\psi(\vz(t))},\label{eq:eulerlagreq}
\end{equation}
where $G_{\overline{\imath}j}$ is the Gram matrix or overlap matrix of the tangent vectors of $\mathcal{M}$:
\begin{equation}
G_{\overline{\imath}j}(\ovz,\vz)=\Braket{\partial_{\overline{\imath}} \psi(\ovz)|\partial_j \psi(\vz)}.\label{eq:gram}
\end{equation}
We can also interpret this as the metric and --- assuming linear independence of the tangent vectors --- define the inverse metric as $G^{i\overline{\jmath}}(\ovz,\vz)$, such that $G^{i\overline{\jmath}}(\ovz,\vz)G_{\overline{\jmath}k}(\ovz,\vz)=\delta^{i}_{\ k}$. The TDVP thus results in
\begin{equation}
\ic\dot{z}^{i}(t) = G^{i\overline{\jmath}}(\ovz(t),\vz(t))\braket{\partial_{\overline{\jmath}} \psi(\ovz(t))|\operator{H}|\psi(\vz(t))}\label{eq:tdvp}
\end{equation}
and its complex conjugate. In the main text, the Euler-Lagrange equations [Eq.~\eqref{eq:eulerlagreq}] were obtained geometrically, by looking for the coefficients $\dot{z}^{i}(t)$ which minimize 
$$\left\lVert \dot{z}^{i}(t)\ket{\partial_i \psi(\vz(t))}-\operator{H}\ket{\psi(\vz(t))}\right\rVert.$$
This minimization is obtained by the orthogonal projection of $\operator{H}\ket{\psi(\vz(t))}$ onto the tangential plane $T_{\vz(t)}\mathcal{M}$, defined by
\begin{equation}
T_{\vz}\mathcal{M}=\mathrm{span}\left\{\Ket{\partial_i \psi(\vz)},i=1,\ldots,n\right\}.\label{eq:tanplane}
\end{equation}
The orthogonal projector $\operator{P}_{T_{\vz}\mathcal{M}}$ is indeed given by
\begin{equation}
\operator{P}_{T_{\vz}\mathcal{M}}(\ovz,\vz)=\ket{\partial_i \psi(\vz)}G^{i\overline{\jmath}}(\ovz,\vz) \bra{\partial_{\overline{\jmath}} \psi(\ovz)},\label{eq:tanplaneproj}
\end{equation}
where the inverse of the Gram matrix appears in order to obtain $\operator{P}_{T_{\vz}\mathcal{M}}^{2}= \operator{P}_{T_{\vz}\mathcal{M}}$.

Whereas Hamiltonian evolution in $\mathcal{H}$ is unitary and thus norm-preserving, this is no longer guaranteed for the projected evolution. In order to ensure norm preservation, we can define a modified Lagrangian $\widetilde{L}(\overline{\psi}(t),\psi(t))=L(\overline{\psi}(t),\psi(t))/\braket{\psi(t)|\psi(t)}$, which results in
\begin{equation}
\widetilde{L}(\ovz(t),\vz(t))=\frac{\ic}{2}\left(\dot{z}^{j}(t)\partial_{j}-\dot{\oz}^{\overline{\jmath}}(t)\partial_{\overline{\jmath}}\right)\ln N(\ovz(t),\vz(t))-H(\ovz(t),\vz(t))\label{eq:modlagrangian}
\end{equation}
where
\begin{align*}
N(\ovz,\vz)&=\braket{\psi(\ovz)|\psi(\vz)},&H(\ovz,\vz)&=\frac{\braket{\psi(\ovz)|\operator{H}|\psi(\vz)}}{\braket{\psi(\ovz)|\psi(\vz)}}.
\end{align*}
Stationarity under variations $\ovz(t)+\overline{\delta \vz}(t)$ now results in the Euler-Lagrange equations
\begin{equation}
\ic \widetilde{G}_{\overline{\imath}j}(\ovz(t),\vz(t))\dot{z}^{j}(t)=H_{\overline{\imath}}(\ovz(t),\vz(t)), \label{eq:modeulerlagreq}
\end{equation}
and complex conjugates, where we have introduced the modified Gram matrix
\begin{equation}
\begin{split}
\widetilde{G}_{\overline{\imath}j}(\ovz,\vz)=\partial_{\overline{\imath}}\partial_{j}\ln N(\ovz,\vz)&=\frac{G_{\overline{\imath}j}(\ovz,\vz)}{N(\ovz,\vz)}-\frac{\braket{\partial_{\overline{\imath}}\psi(\ovz)|\psi(\vz)}\braket{\psi(\ovz)|\partial_{j}\psi(\vz)}}{N(\ovz,\vz)^{2}}\\
&=\frac{\braket{\partial_{\overline{\imath}}\psi(\ovz)|\partial_{j}\psi(\vz)}}{\braket{\psi(\ovz)|\psi(\vz)}}-\frac{\braket{\partial_{\overline{\imath}}\psi(\ovz)|\psi(\vz)}\braket{\psi(\ovz)|\partial_{j}\psi(\vz)}}{\braket{\psi(\ovz)|\psi(\vz)} ^{2}},
\end{split}
\end{equation}
and the gradient of the normalized expectation value
\begin{displaymath}
H_{\overline{\imath}}(\ovz,\vz)=\partial_{\overline{\imath}}H(\ovz,\vz)=\frac{\braket{\partial_{\overline{\imath}}\psi(\ovz)|\operator{H}|\psi(\vz)}}{N(\ovz,\vz)}-\frac{\braket{\partial_{\overline{\imath}}\psi(\ovz)|\psi(\vz)}}{N(\ovz,\vz)}H(\ovz,\vz)=\frac{\braket{\partial_{\overline{\imath}}\psi(\ovz)|\operator{H}|\psi(\vz)}}{\braket{\psi(\ovz)|\psi(\vz)}}-\frac{\braket{\partial_{\overline{\imath}}\psi(\ovz)|\psi(\vz)}\braket{\psi(\ovz)|\operator{H}|\psi(\vz)}}{\braket{\psi(\ovz)|\psi(\vz)}^{2}}.\label{eq:modgrad}
\end{displaymath}
These expressions can be easily interpreted. Under an infinitesimal variation, the norm or phase of a state $\ket{\psi}$  changes if we move in the direction of $\ket{\psi}$. Norm conservation is thus obtained when we subtract from every tangent vector $\ket{\partial_{i}\psi(\vz)}$ its component along $\ket{\psi(\vz)}$ by replacing it with $\operator{P}_{0}(\ovz,\vz)\ket{\partial_{i}\psi(\vz)}$, where the projector $\operator{P}_{0}$ is given by
\begin{equation}
\operator{P}_{0}(\ovz,\vz)=\operator{1}-\frac{\ket{\psi(\vz)}\bra{\psi(\ovz)}}{\braket{\psi(\ovz)|\psi(\vz)}},
\end{equation}
and thus $\operator{P}_{0}(\ovz,\vz)\ket{\partial_{i}\psi(\vz)}=\ket{\partial_{i}\psi(\vz)}-\ket{\psi(\vz)} N(\ovz,\vz)^{-1} \braket{\psi(\ovz)|\partial_{i}\psi(\vz)}$. We indeed find
\begin{displaymath}
\tilde{G}_{\overline{\imath}j}(\ovz,\vz)=N(\ovz,\vz)^{-1} \braket{\partial_{\overline{\imath}}\psi(\ovz)|\operator{P}_{0}(\ovz,\vz) | \partial_{j}\psi(\vz)}
\end{displaymath}
and
\begin{displaymath}
H_{\overline{\imath}}(\ovz,\vz)=N(\ovz,\vz)^{-1} \braket{\partial_{\overline{\imath}}\psi(\ovz)|\operator{P}_{0}(\ovz,\vz) \operator{H}|\psi(\vz)}.
\end{displaymath}

If the manifold $\mathcal{M}$ allows for norm and phase variations of states, \textit{i.e.}\ if $\ket{\psi}\in T\mathcal{M}$, then we can define the contravariant vector $\psi^{i}$ such that $\psi^{i}\ket{\partial_{i}\psi}=\ket{\psi}$. In this paragraph, we henceforth omit all arguments $\vz$ and $\ovz$ for the sake of brevity. By definition we have that $\operator{P}_{0}\ket{\partial_{i}\psi}\psi^{i}=0$ and we can conclude that $\widetilde{G}_{\overline{\imath}j}$ has an eigenvalue zero, since $\widetilde{G}_{\overline{\imath}j}\psi^{j}=0=\overline{\psi}^{\overline{\imath}}G_{\overline{\imath}j}$, from which we immediately obtain the corresponding eigenvector. (Note that $G_{\overline{\imath}j}$ is a Hermitian matrix.) We now also define the covariant vector $\psi_{\overline{\imath}}=G_{\overline{\imath}j}\psi^{j}=\braket{\partial_{\overline{\imath}} \psi|\psi}$ so that $\overline{\psi}_{i}\psi^{i}=\braket{\psi|\psi}=N$. With these definitions, we can write $\widetilde{G}_{\overline{\imath}j}=N^{-1} G_{\overline{\imath}j}-N^{-2}\psi_{\overline{\imath}}\overline{\psi}_{j}$. Even though $\widetilde{G}_{\overline{\imath}j}$ is not invertible, we can still define a pseudo-inverse as $\widetilde{G}^{i\overline{\jmath}}=N G^{i\overline{\jmath}}-\psi^{i}\overline{\psi}^{\overline{\jmath}}$, so that $\widetilde{G}^{i\overline{\jmath}} \widetilde{G}_{\overline{\jmath}k}=\delta^{i}_{\ k}-N^{-1}\psi^{i}\overline{\psi}_{k}$ and $\widetilde{G}_{\overline{\imath}j} \widetilde{G}^{j\overline{k}}=\delta^{\ \overline{k}}_{\overline{\imath}}-N^{-1}\psi_{\overline{\imath}}\overline{\psi}^{\overline{k}}$. Since we can rewrite $H_{\overline{\imath}}$ as $N^{-1} (\delta^{\ \overline{k}}_{\overline{\imath}}-N^{-1}\psi_{\overline{\imath}}\overline{\psi}^{\overline{k}}) \braket{\partial_{\overline{k}}\psi|\operator{H}|\psi}$, we are allowed to apply this pseudo-inverse to the Euler-Lagrange equations in order to obtain
\begin{equation}
\ic \dot{z}^{i}(t)= \widetilde{G}^{i\overline{\jmath}}(\ovz(t),\vz(t))  H_{\overline{\jmath}}(\ovz(t),\vz(t)).
\end{equation}
In principle, the component of $\dot{\vz}(t)$ along the zero eigenspace of $\widetilde{G}$ can be chosen freely but, with the particular solution in the equation above, we satisfy $\overline{\psi}_{i}\dot{z}^{i}(t)=\braket{\psi|\partial_{i}\psi}\dot{z}^{i}(t)=0$, which is the required condition for norm (and phase) conservation.

If the manifold $\mathcal{M}$ does not contain the freedom to change the norm and phase of a state, the modified metric $\widetilde{G}$ would have the same rank as the original metric $G$, which we assumed to be invertible. In particular, if $\ket{\psi(\vz)}\perp T_{\vz}\mathcal{M}$, then $\widetilde{G}_{\overline{\imath},j}(\ovz,\vz)=N(\ovz,\vz)^{-1} G_{\overline{\imath},j}(\ovz,\vz)$ and $H_{\overline{\imath}}(\ovz,\vz)= N(\ovz,\vz)^{-1} \braket{\partial_{\overline{\imath}}\psi(\ovz)|\operator{H}|\psi(\vz)}$. The Euler-Lagrange equations following from $S\{\ovz(t),\vz(t)\}$ or from $\widetilde{S}\{\ovz(t),\vz(t)\}$ are then identical. 

Finally, by defining for every pair of functions $f(\ovz,\vz)$ and $g(\ovz,\vz)$ a Poisson bracket
\begin{equation}
\{f,g\}=\partial_{i}f \widetilde{G}^{i\overline{\jmath}}\partial_{\overline{\jmath}}g-\partial_{i}g \widetilde{G}^{i\overline{\jmath}}\partial_{\overline{\jmath}}f
\end{equation}
we can write down the Euler-Lagrange equations as
\begin{align}
\dot{z}^{i}&=\ic \{H,z^{i}\},&\dot{\oz}^{\overline{\imath}}&=\ic \{H,\oz^{\overline{\imath}}\}.
\end{align}
For every operator $\operator{O}$ acting on $\mathcal{H}$, we can define the expectation value $O(\ovz,\vz)=\braket{\psi(\ovz)|\operator{O}|\psi(\vz)}/\braket{\psi(\ovz)|\psi(\vz)}$ so that its time evolution is governed by $\dot{O}=\ic\{H,O\}$. The manifold $\mathcal{M}$ is thus a symplectic manifold. From the antisymmetry of the Poisson bracket we find $\{H,H\}=0$, which implies that the energy of the state $\ket{\psi}\in\mathcal{M}$ is conserved under exact integration of the TDVP equations.

The symplectic properties of the time-dependent variational principle also conserve other symmetries. Assume that the Hamiltonian is invariant under the action of a symmetry operator $\operator{U}$ (which should be a unitary operator), such that $[\operator{H},\operator{U}]=0$. In order to be able to transfer this symmetry to the manifold $\mathcal{M}$, we need to assume that for any state $\ket{\psi(\vz)}\in\mathcal{M}$, the action of $\operator{U}$ is mapped to a new state $\ket{\psi(\vu(\vz))}=\operator{U}\ket{\psi(\vz)}\in\mathcal{M}$. Because of the unitarity of $\operator{U}$, we have $N(\ovu(\ovz),\vu(\vz))= N(\ovz,\vz)$, from which we obtain
\begin{equation}
\partial_{\overline{\imath}}\ou^{\overline{\jmath}}(\ovz) \widetilde{G}_{\overline{\jmath},k}(\ovu(\ovz),\vu(\vz)) \partial_l u^k(\vz)= G_{\overline{\imath},l}(\ovz,\vz),
\end{equation}
The condition $[\operator{H},\operator{U}]=0$ also allows to conclude $H(\ovu(\ovz),\vu(\vz))=H(\ovz,\vz)$ and thus
\begin{equation}
\begin{split}
&\partial_{\overline{\imath}}\ou^{\overline{\jmath}}(\ovz) H_{\overline{\jmath}}(\ovu(\ovz),\vu(\vz))=H_{\overline{\imath}}(\ovz,\vz),\\
&H_{j}(\ovu(\ovz),\vu(\vz))\partial_{i}u^{j}(\vz) =H_{i}(\ovz,\vz).
\end{split}
\end{equation}
The metric and the gradient thus transform covariantly under the symmetry transformation and can be used to transform Eq.~\eqref{eq:modeulerlagreq} into
\begin{displaymath}
+\ic\partial_{\overline{\imath}}\ou^{\overline{\jmath}}(\ovz(t)) \widetilde{G}_{\overline{\jmath},k}(\ovu(\ovz(t)),\vu(\vz(t))) \frac{\d\ }{\d t}u^k(\vz(t))=\partial_{\overline{\imath}}\ou^{\overline{\jmath}}(\ovz(t)) H_{\overline{\jmath}}(\ovu(\ovz(t)),\vu(\vz(t))),
\end{displaymath}
and its complex conjugate. By using the injectivity of the map $\vu(\vz)$, we can eliminate the Jacobians $\partial_{\overline{\imath}} \ou^{\overline{\jmath}}$ and $\partial_k u^j$ in order to obtain the correct flow equations in terms of the new coordinates $(\vu(t),\ovu(t))$. One case that is not covered by this general derivation is when $\operator{U}$ is an anti-linear operator, since $\vu$ will then depend on $\vz$ anti-holomorphically. Anti-linear transformations appear in quantum mechanics exclusively for time-reversal transformations. Let us denote $\operator{R}\ket{\psi(\vz)}=\ket{\psi(\vec{r}(\ovz))}$ with $\operator{R}$ the operator of an elementary time-reversal transformation. Because of the anti-unitarity of $\operator{R}$ and its commutation relation with the Hamiltonian (\textit{i.e.} $[\operator{H},\operator{R}]=0$ since we assume $\operator{H}$ to be time-reversal invariant), we obtain $N(\overline{\vec{r}}(\vz)),\vec{r}(\ovz))=\overline{N(\ovz,\vz)}=N(\ovz,\vz)$ and $H(\overline{\vec{r}}(\vz),\vec{r}(\ovz))=\overline{H(\ovz,\vz)}=H(\ovz,\vz)$, which yields
\begin{equation}
\partial_{i}\overline{r}^{\overline{\jmath}}(\vz) \widetilde{G}_{\overline{\jmath},k}(\overline{\vec{r}}(\vz),\vec{r}(\ovz)) \partial_{\overline{l}} r^k(\ovz)= \widetilde{G}_{\overline{l},i}(\ovz,\vz),
\end{equation}
and
\begin{equation}
\begin{split}
&\partial_{i}\overline{r}^{\overline{\jmath}}(\vz) H_{\overline{\jmath}}(\overline{\vec{r}}(\vz),\vec{r}(\ovz))=H_{i}(\ovz,\vz),\\
& H_{j}(\overline{\vec{r}}(\vz),\vec{r}(\ovz))\partial_{\overline{\imath}}r^{j}(\ovz)=H_{\overline{\imath}}(\ovz,\vz).
\end{split}
\end{equation}
These relations convert Eq.~\eqref{eq:modeulerlagreq} into
\begin{displaymath}
\begin{cases}
+\ic\frac{\d\ }{\d t}\overline{r}^{\overline{\jmath}}(\vz(t)) \widetilde{G}_{\overline{\jmath},k}(\overline{\vec{r}}(\vz(t)),\vec{r}(\ovz(t))) \partial_{\overline{\imath}} r^k(\ovz(t)) = H_{k}(\overline{\vec{r}}(\vz(t)),\vec{r}(\ovz(t)))\partial_{\overline{\imath}}r^{k}(\ovz(t)),\\
-\ic \partial_i\overline{r}^{\overline{k}}(\vz(t)) \widetilde{G}_{\overline{k},j}(\overline{\vec{r}}(\vz(t)),\vec{r}(\ovz(t)))\frac{\d\ }{\d t} r^{j}(\ovz(t)) = \partial_{i}\overline{r}^{\overline{k}}(\vz(t)) H_{\overline{k}}(\overline{\vec{r}}(\vz(t)),\vec{r}(\ovz(t))),
\end{cases}
\end{displaymath}
or, by eliminating the Jacobian of the transformation,
\begin{equation}
\begin{cases}
-\ic \widetilde{G}_{\overline{\imath},j}(\overline{\vec{r}}(\vz(t)),\vec{r}(\ovz(t)))\frac{\d\ }{\d t} r^{j}(\ovz(t)) =  H_{\overline{\imath}}(\overline{\vec{r}}(\vz(t)),\vec{r}(\ovz(t))),\\
+\ic\frac{\d\ }{\d t}\overline{r}^{\overline{\jmath}}(\vz(t)) \widetilde{G}_{\overline{\jmath},i}(\overline{\vec{r}}(\vz(t)),\vec{r}(\ovz(t))) = H_{i}(\overline{\vec{r}}(\vz(t)),\vec{r}(\ovz(t)))
\end{cases}
\end{equation}
Note that the signs of the two equations have been switched, which is necessary to revert the time evolution of the new coordinates $(\overline{\vec{r}}(t),\vec{r}(t))$. For a time-reversal invariant Hamiltonian $\operator{H}$ and a variational manifold $\mathcal{M}$ that contains the time-reversed state $\operator{R}\ket{\psi(\vz)}\in\mathcal{M}$ for each of its elements $\ket{\psi(\vz)}\in\mathcal{M}$, the flow equations of the time-dependent variational principle are also time-reversal invariant. This can be exploited in numerical integration schemes in order to construct symmetric schemes with improved stability (see next section).

Returning to the case of linear symmetry transformations $\operator{U}$, the expectation value $U(\ovz,\vz)$ is a constant of motion of the exact evolution according to the TDSE. For a general (discrete or continuous) symmetry, even when $\operator{U}\ket{\psi(\vz)}=\ket{\psi(\vu(\vz))}\in \mathcal{M}$, we do not have that $\operator{U}\ket{\psi(\vz)}\in T_{\vz}\mathcal{M}$. Consequently, $U(\ovz(t),\vz(t))$ is not automatically a constant of motion of the TDVP evolution. But often $U(\ovz,\vz)$ does not represent an interesting quantity. However, when the symmetry operator $\operator{U}$ corresponds to a continuous symmetry generated by the Hermitian generator $\operator{K}$, with $[\operator{K},\operator{H}]=0$, then the expectation value of $\operator{K}$ is an interesting constant of motion for the exact evolution according to the TDSE. We define a one-parameter family of transformations $\operator{U}(\epsilon)=\exp(\ic \epsilon \operator{K})$. Since we require that for every state $\ket{\psi(\vz)}$ in the manifold $\mathcal{M}$, $\operator{U}(\epsilon)\ket{\psi(\vz)}=\ket{\psi(\vu(\vz,\epsilon))}\in\mathcal{M}$, we can differentiate this defining relation with respect to $\epsilon$ and set $\epsilon=0$ in order to learn
\begin{equation}
\ic \operator{K} \ket{\psi(\vz)}= \frac{\partial u^i}{\partial \epsilon}(\vz,0)\ket{\partial_{i}\psi(\vz)}=k^{i}(\vz)\ket{\partial_{i}\psi(\vz)}.
\end{equation}
The action of $\operator{K}$ on a state $\ket{\psi(\vz)}$ thus has to be exactly captured in $T_{\vz}\mathcal{M}$. We obtain for the part orthogonal to $\ket{\psi(\vz)}$
\begin{align*}
\braket{\partial_{\overline{\jmath}}\psi(\ovz)|\operator{P}_{0}(\ovz,\vz)|\partial_{i}\psi(\vz)} k^{i}(\vz)&= \ic\braket{\partial_{\overline{\jmath}} \psi(\ovz)|\operator{P}_{0}(\ovz,\vz)\operator{K}|\psi(\vz)}=\ic\braket{\partial_{\overline{\jmath}} \psi(\ovz)|\operator{K}|\psi(\vz)}-\ic\braket{\partial_{\overline{\jmath}}\psi(\ovz)|\psi(\vz)}K(\ovz,\vz)\\
\Leftrightarrow \qquad N(\ovz,\vz)\widetilde{G}_{\overline{\jmath},i}(\ovz,\vz)k^{i}(\vz)&=\ic N(\ovz,\vz) \partial_{\overline{\jmath}} K(\ovz,\vz)
\end{align*}
and for the part parallel to $\ket{\psi(\vz)}$
\begin{align*}
\braket{\psi(\ovz)|\partial_{i}\psi(\vz)} k^{i}(\vz)&= \ic\braket{\psi(\ovz)|\operator{K}|\psi(\vz)}\\
\Leftrightarrow \qquad \overline{\psi}_{i}(\ovz,\vz) k^{i}(\vz)&=\ic N(\ovz,\vz) K(\ovz,\vz).
\end{align*}
The combination of these two equations results in:
\begin{equation}
k^{i}(\vz)= \ic \widetilde{G}^{i,\overline{\jmath}}(\ovz,\vz) \partial_{\overline{\jmath}} K(\ovz,\vz)+\ic K(\ovz,\vz) \psi^{i}(\vz).
\end{equation}
If we now express $H(\ovz+\varepsilon \overline{\vec{k}}(\ovz), \vz+\varepsilon\vec{k}(\vz))=H(\ovz,\vz)$ to first order in $\varepsilon$, we obtain from the first-order expansion
\begin{displaymath}
\ic\varepsilon\left(\partial_{i} H \widetilde{G}^{i,\overline{\jmath}}\partial_{\overline{\jmath}} K-\partial_{i} K \widetilde{G}^{i,\overline{\jmath}} \partial_{\overline{\jmath}} H + \partial_{i}H \psi^{i} K- K \overline{\psi}^{\overline{\jmath}} \partial_{\overline{\jmath}} H\right)=0
\end{displaymath}
(where we have omitted the arguments for the sake of simplicity) and thus, using $\psi^{i}\partial_{i} H = \overline{\psi}^{\overline{\jmath}}\partial_{\overline{\jmath}} H = H$, we find
\begin{equation}
\{K,H\} = 0,
\end{equation}
so that any generator of a symmetry transformation of the Hamiltonian which can be exactly captured within the variational manifold produces a constant of motion $K(\ovz(t),\vz(t))=K(\ovz(0),\vz(0))$.

Finally, we can use the TDVP to simulate imaginary time evolution by setting $t=-\ic \tau$. Imaginary time evolution according to the (imaginary) TDSE equation in the full Hilbert space $\mathcal{H}$ produces a gradient descent which converges to the true ground state $\ket{\psi_{0}}$ for any initial state $\ket{\psi}$ such that $\braket{\psi_{0}|\psi} \neq 0$. There are no local minima in which the gradient descent can be trapped. Excited states correspond to saddle points. The imaginary TDVP evolution is the best approximation to this gradient descent in $\mathcal{H}$; it is set apart from a gradient descent in parameter space (with $H_{i}=\partial_{i}$ the gradient) by the appearance of the Gram matrix. Only when the Gram matrix $\tilde{G}$ is the unit matrix, is the imaginary TDVP evolution equal to a gradient descent in parameter space. For the TDVP flow, the time derivative of the energy expectation value is given by
\begin{equation}
\frac{\d }{\d \tau} H(\ovz(\tau),\vz(\tau))=-2 \partial_{i} H \widetilde{G}^{i,\overline{\jmath}}\partial_{\overline{\jmath}} H \leq 0.
\end{equation}
The energy expectation value thus monotonically decreases until we reach the minimum, which is characterized by $\partial_{i} H(\ovz,\vz)= \partial_{\overline{\jmath}} H(\ovz,\vz)$. Note that if we define the norm of the gradient as $\eta=[\partial_{i} H \widetilde{G}^{i,\overline{\jmath}}\partial_{\overline{\jmath}} H ]^{1/2}$, then the rate of change of the energy expectation value is $\mathcal{O}(\eta^{2})$. The energy thus converges quadratically faster than the state itself, which is a well-known result. 

Clearly, this whole section is applicable to the case of MPS and produces the results of the main text. In particular, the action of all symmetries for which the generator can be written as a sum of one-site terms can be exactly captured by the variational manifold $\mathcal{M}_{\text{uMPS}}$ and thus produces constants of motion. The only difficulty in the MPS representation is its gauge invariance, which can be interpreted as an overparameterization and thus results in a set of tangent vectors which are not all linearly independent. But we have discussed in the main text how to overcome this difficulty. Since $\ket{\partial_{i}\psi} B_{X}^{i}$ is identically zero for all tangent vectors which consist of pure gauge transforms, any overlap or expectation value is also zero in the corresponding block. By using the linear representation $B^{i}(x)$, we are effective projecting all relevant quantities into their non-zero subspace, which is $(d-1)D^{2}$ dimensional.

\section{Details of the implementation of the TDVP for uMPS}
In this section we discuss additional details of the implementation of the TDVP for uMPS. 

\subsection{Choice of gauge}
In the main text we have discussed how to fix the gauge of the variations $B^{i}$ describing states in the tangent plane $T_{A}\mathcal{M}_{\text{uMPS}}$. We call this gauge fixing constraint the \emph{left gauge-fixing condition}, since it ensures that the left eigenvector and the largest eigenvalue (thus norm-preserving) of the transfer operator $E^{A}_{A}$ do not change to first order. Similar results could have been obtained with the following \emph{right gauge-fixing condition}. We define the $dD\times D$ matrix
\begin{equation}
R_{(\alpha s),\beta}=[r^{1/2}(A^{s})^{\dagger}]_{\alpha\beta}
\end{equation}
and define the $(d-1)D\times dD$ matrix $V_{R}$ that satisfies $V_{R} R=0$ and $V_{R} V_{R}^{\dagger}=1$. Thus, $V_{R}^{\dagger}$ contains an orthonormal basis for the null space of $R^{\dagger}$. If we now also introduce  $[V^{s}_{R}]_{\gamma,\beta}=[V_{R}]_{\gamma,(s\beta)}$ (with $\gamma=1,\ldots,(d-1)D$, $s=1,\ldots,d$, and $\beta=1,\ldots,D$) then we find a different parameterization of the tangent plane as 
\begin{equation}
\widetilde{B}^{s}(x)=l^{-1/2} x V^{s}_{R} r^{-1/2},
\end{equation}
where the independent parameters $x$ have now been grouped in a $D\times (d-1)D$ matrix. This parameterization satisfies the right gauge-fixing constraint $E^{\widetilde{B}(x)}_{A}\rket{r}=0$ and also produces $\overline{\widetilde{B}}^{\overline{\imath}}(x) G_{\overline{\imath}j}\widetilde{B}^{j}(y) = |\mathbb{Z}| \mathrm{tr}[x^{\dagger} y]$.

Up to this point, we have not elaborated on convenient gauges for $A$ itself. A left orthonormalization gauge --- where $\sum_{s=1}^{d} (A^{s})^{\dagger} A^{s}=1$ and thus $l=1$ --- combines very well with the first-order conservation of the left eigenvector by variations $B$. Similarly, a right orthonormalization gauge --- where $r=1$ --- combines very well with the right gauge fixing condition on $B$. However, both choices have also a numerical disadvantage. The representations $B(x)$ and $\widetilde{B}(x)$ rely heavily on calculating $l^{-1/2}$ and $r^{-1/2}$, which both appear in the different terms in the relevant expressions but never simultaneously (if worked out correctly). Hence, we would like to condition both matrices equally well at the same time. But if, \textit{e.g.}, $l=1$, then $r$ contains the eigenvalues of the density matrix of half of the chain, which are the square of Schmidt coefficients. If we try to obtain an accurate approximation where the smallest eigenvalues are of the order of machine precision, $r$ is not determined accurately, it is ill-conditioned and the many operations with $r^{-1/2}$ can produce large numerical errors. Consequently, we have chosen to fix the gauge of $A$ by requiring that $l=r=\lambda$ where $\lambda$ is the diagonal matrix containing the Schmidt coefficients. Starting from a general $A$, we  first find the canonical form $(\lambda,\Gamma)$ [see G.~Vidal, {Phys. Rev. Lett.} {\bf 98}, 070201 (2007)] and then set $A^{s}=\lambda^{1/2} \Gamma^{s} \lambda^{1/2}$. This choice of gauge, which we  call the \emph{symmetric gauge}, evenly distributes the small eigenvalues of the density matrix over the left and the right eigenvector of the transfer operator, resulting in a better conditioned algorithm. Nevertheless, these stability considerations are still the main reason that we cannot converge states up to machine precision (\textit{i.e.}\ $\eta\approx10^{-15}$) at larger values of $D$ where the spectrum of Schmidt coefficients contains many small values. 

\subsection{Measures of convergence and error for uMPS}
Let's now reinterpret the TDVP evolution in order to assess both the convergence and the error of the TDVP equations. A measure of convergence is of course only useful for imaginary time evolution where we expect to end up in a steady state which is presumably the global minimum of the energy function $H(\overline{A},A)$ and thus the best possible approximation of the ground state of $\operator{H}$ in the variational manifold $\mathcal{M}_{\text{uMPS}}$. Norm conservation is obtained by replacing the evolution vector $\operator{H}\ket{\psi}$ of the Schr\"odinger equation by
\begin{equation}
\operator{P}_{0}(\overline{A},A)\operator{H}\ket{\psi(A)}=(\operator{1}-\ket{\psi(A)}\bra{\psi(\overline{A})})\operator{H}\ket{\psi(A)}=(\operator{H}-\braket{\psi(\overline{A})|\operator{H}|\psi(A)})\ket{\psi}=(\operator{H}-H(\overline{A},A))\ket{\psi(A)},
\end{equation}
where we have assumed that $\ket{\psi(A)}$ is normalized to one. For a translation-invariant Hamiltonian $\operator{H}=\sum_{n\in \mathbb{Z}} \operator{T}^{n}\operator{h}\operator{T}^{-n}$, we introduce the notation $\braket{\psi(\overline{A})|\operator{H}|\psi(A)}=H(\overline{A},A)=|\mathbb{Z}| h(\overline{A},A)$, where $h(\overline{A},A)=\braket{\psi(\overline{A})|\operator{h}|\psi(A)}$.  Once again, we  henceforth omit all arguments. If $\operator{H}$ is a nearest-neighbor Hamiltonian, so that $\operator{h}$ only acts nontrivially on sites zero and one, we have $h=\rbraket{l|H^{AA}_{AA}|r}$ and we can write the exact evolution vector as
\begin{equation}
(\operator{H}-H)\ket{\psi}=\sum_{n\in\mathbb{Z}} \operator{T}^{n} v_{\mathrm{L}}^{\dagger}\left(\cdots A^{s_{-2}}A^{s_{-1}} C^{s_{0}s_{1}}A^{s_{2}}\cdots\right)v_{\mathrm{R}}\ket{\ldots s_{-2}s_{-1}s_{0}s_{1}s_{2}\ldots}\label{eq:exacthpsi}
\end{equation}
with $C^{st}=\sum_{u,v=1}^{d}\braket{s,t|\operator{h}-h|u,v} A^{u}A^{v}$. The TDVP equation boils down to projecting this vector into the tangent plane $T_{A}\mathcal{M}_\text{uMPS}$, and thus finding the set of coefficients $x^{\ast}$ such that
\begin{equation}
B^{i}(x^{\ast})\ket{\partial_{i}\psi}=\sum_{n\in\mathbb{Z}}\operator{T}^{n} \left(\sum_{\{s_{j}\}=1}^{d}v_{\mathrm{L}}^{\dagger}( \cdots A^{s_{-1}}B^{s_{0}}(x^{\ast})A^{s_{+1}}\cdots)v_{\mathrm{R}} \ket{\ldots s_{-1}s_{0}s_{+1}\ldots}\right)
\end{equation}
minimizes $\lVert B^{i}(x)\ket{\partial_{i}\psi} - (\operator{H}-H)\ket{\psi}\rVert$. An algorithm for efficiently determining the set of the coefficients $x^{\ast}$ was explained in the main text. The additional term $H\ket{\psi}$, not present in the main text, is non-essential, as it produces norm preservation which is automatically satisfied since $\braket{\psi|\partial_{i}\psi} B^{i}(x)=0$ by construction. We now formally set
\begin{displaymath}
B^{i}(x^{\ast})\ket{\partial_{i}\psi}=\operator{P}_{T\mathcal{M}}(\operator{H}-H)\ket{\psi}
\end{displaymath}
as the vector that describes the TDVP flow.

Let us now further elaborate on the properties of the TDVP flow for imaginary time evolution with uMPS. For imaginary time evolution, we can use the general result from the previous section to write
\begin{multline}
\frac{\d }{\d \tau} h(\overline{A}(\tau),A(\tau))=\frac{1}{|\mathbb{Z}|} \frac{\d }{\d \tau} H(\overline{A}(\tau),A(\tau))=-\frac{1}{|\mathbb{Z}|} \lVert \operator{P}_{T\mathcal{M}}(\operator{H}-H)\ket{\psi} \rVert^{2}\\=-\frac{1}{|\mathbb{Z}|} \lVert B^{i}(x^{\ast})\ket{\partial_{i}\psi} \rVert^{2}=-\mathrm{tr}\left[(x^{\ast})^{\dagger}(x^{\ast})\right].
\end{multline}
We obtain here some peculiar properties of quantum states of systems with infinite size (\textit{i.e.}\ in the thermodynamic limit). Since the rate of change of the state (\textit{i.e.}\ the norm of the gradient $\lVert B^{i}(x^{\ast})\ket{\partial_{i}\psi}\rVert$) is proportional to $|\mathbb{Z}|^{1/2}$, an infinite size uMPS  always has zero overlap with the state it is converging to and with every other non-equivalent uMPS. This phenomenon is true in general for translation invariant states of a system of infinite size and was dubbed the (infrared) orthogonality catastrophe by Anderson [P.~W.~Anderson, Phys. Rev. Lett. {\bf 18}, 1049 (1967)]. Indeed, let $\ket{\psi(A)}$ and $\ket{\psi(\widetilde{A})}$ be two normalized uMPSs (which requires that the largest eigenvalue of $E^{A}_{A}$ and of $E^{\widetilde{A}}_{\widetilde{A}}$ is $1$), where we also assume that both uMPS have a transfer matrix with a unique eigenvalue with magnitude one. We can then then compute the spectral radius $\rho(E^{\tilde{A}}_{A})$ of $E^{\tilde{A}}_{A}$, which is also know as the ground state fidelity per site $d(\overline{A},\tilde{A})$, since the total fidelity between the two quantum states scales as $|\braket{\psi(\overline{A})|\psi(\tilde{A})}|=F(\overline{A},\tilde{A})=d(\overline{A},\tilde{A})^{|\mathbb{Z}|}$. Then either $d(\overline{A},\tilde{A})<1$, such that $\braket{\psi(\widetilde{A})|\psi(A)}=0$, or $d(\overline{A},\tilde{A})=1$, in which case we can prove the existence of $\phi\in [0,2\pi)$ and $G\in \mathsf{GL}(D)$ such that $A^{s}= \mathrm{e}^{\ic \phi} G \widetilde{A}^{s} G^{-1}$ and $\ket{\psi(A)} \sim \ket{\psi(\widetilde{A})}$, provided that the boundary vectors $v_{\mathrm{L}}$ and $v_{\mathrm{R}}$ are chosen identically. Two non-equivalent infinite size states ($d(\overline{\tilde{A}},A) < 1$) can however be locally similar and it is useful to define the \emph{local} rate of change on the state as $\eta=\sqrt{\mathrm{tr}[(x^{\ast})^{\dagger}x^{\ast}]}$. We then obtain for the rate of change of the energy density $dh/dt=-\eta^{2}$, which reproduces the quadratically faster convergence of the energy expectation value. Since convergence of the energy expectation value is the basis of all algorithms based on the (time-independent) variational principle, the corresponding state can still be far from the theoretically optimal state in the variational manifold. The same error  propagates to all other expectation values of local operators. But in this imaginary time algorithm we can use the local change  $\eta$ of the state as a convergence measure. At the point where the expectation value of the energy density is already at machine precision, we can further evolve the state in order to decrease $\eta$ to near machine precision and thus to very closely approach the theoretically optimal uMPS. Of course, the maximal accuracy that can be obtained  depends on the conditioning of $l$ and $r$, which are used in the construction of $x^{\ast}$ and $B (x^{\ast})$, as discussed in the previous subsection. 

On the other hand, we can also assess the error between the theoretically optimal uMPS and the exact ground state. In fact, we can measure the error we make at any point in the TDVP evolution, both for real and imaginary time evolution. Note that the exact evolution vector is written in Eq.~\eqref{eq:exacthpsi} in a form that is very similar to the tangent vectors, except for the fact that it acts non-trivially on two sites at the same time. By projecting onto the tangent plane, we are missing a part $(\operator{1}-\operator{P}_{T\mathcal{M}_{\text{uMPS}}})(\operator{H}-H)\ket{\psi}$ of the exact evolution. We can thus assess the tendency of the exact evolution $(\operator{H}-H)\ket{\psi}$ to take $\ket{\psi}$ out of the manifold $\mathcal{M}_{\text{uMPS}}$ by calculating
\begin{equation}
\begin{split}
\lVert(\operator{1}-\operator{P}_{T\mathcal{M}_{\text{uMPS}}})(\operator{H}-H)\ket{\psi}\rVert&=\sqrt{\Delta H^{2}-\overline{B}^{\overline{\imath}}(\overline{x}^{\ast})\partial_{\overline{\imath}}H-\partial_{i}H B^{i}(x^{\ast})+\overline{B}^{\imath}(\overline{x}^{\ast}) \widetilde{G}_{\overline{\imath}j}B^{j}(x^{\ast})}\\
&=\sqrt{\Delta H^{2}-\overline{B}^{\overline{\imath}}(x^{\ast}) \braket{\partial_{\overline{\imath}} \psi|\partial_{j} \psi} B^{j}(x^{\ast})}\\
&=\sqrt{\Delta H^{2}-\partial_{i} H G^{i\overline{\jmath}} \partial_{\overline{\jmath}}H}.\label{eq:globalerror}
\end{split}
\end{equation}
If this quantity is zero, then the evolution is exactly captured within our manifold, and the Hamiltonian can be thought of as effectively acting on a single site. In the final optimum where $\lVert B^{i}(x^{\ast}) \ket{\partial_{i} \psi} \rVert = 0$, the global first order error on the state is then given by the familiar expression $\lVert (\operator{H}-H)\ket{\psi}\rVert=\Delta H$, which we now evaluate to find
\begin{displaymath}
\Delta H^{2}=\braket{\psi|(\operator{H}-H)^{2}|\psi}=\sum_{n,m\in\mathbb{Z}}\braket{\psi|\operator{T}^{n}(\operator{h}-h)\operator{T}^{m-n}(\operator{h}-h)\operator{T}^{-m}|\psi}.
\end{displaymath}
With $\operator{H}$ being our nearest neighbor Hamiltonian and $\ket{\psi}$ being our translationally invariant uMPS, we obtain
\begin{equation}
\Delta H^{2} =|\mathbb{Z}|\left(\sum_{n=-1}^{1}\braket{\psi|(\operator{h}-h)\operator{T}^{n}(\operator{h}-h)|\psi}
+2\rbraket{l|E^{C}_{AA}(1-E)^{-1}E^{AA}_{C}|r}\right).\label{eq:dh}
\end{equation}
The first term results in
\begin{align*}
\braket{\psi|(\operator{h}-h)(\operator{h}-h)|\psi} &= \sum_{s,t,u,v=1}^{d}\braket{u,v|(\operator{h}-h)^{2}|s,t}\rbraket{l|A^{s}A^{t}\otimes \overline{A}^{u} \overline{A}^{v}|r},\\
\braket{\psi|(\operator{h}-h)\operator{T}(\operator{h}-h)|\psi} &= \sum_{r,s,t,u,v,w=1}^{d}\braket{u,v,w|(\operator{h}-h)\operator{T}(\operator{h}-h) \operator{T}^{-1}|r,s,t}\rbraket{l|A^{r}A^{s}A^{t}\otimes \overline{A}^{u} \overline{A}^{v}\overline{A}^{w}|r},
\end{align*}
for $n=0$ and $n=1$ respectively, and in the complex conjugate of the last expression for $n=-1$. Note that both quantities in the square root in Eq.~\ref{eq:globalerror} are proportional to $|\mathbb{Z}|$. At any point in the evolution, we can thus define a local error on the state as $\epsilon=|\mathbb{Z}|^{-1/2} \lVert (\operator{1}-\operator{P}_{T\mathcal{M}_{\text{uMPS}}})(\operator{H}-H)\ket{\psi}\rVert$. In imaginary time evolution, we converge towards the minimum where $\partial_{i}H=\partial_{\overline{\jmath}}H=0$. We thus find $\epsilon=\sqrt{\Delta H^{2}/|\mathbb{Z}|}$ in the optimal uMPS. The error in the expectation value of local operators is proportional to $\epsilon$, whereas the error in the expectation value of the energy density is proportional to $\epsilon^{2}$. If at some point the error $\epsilon$ exceeds a certain tolerance level we can expand the variational manifold by increasing the MPS bond dimension $D$. An optimal construction for this is described in the next section. This same construction also yields a better way to evaluate $\lVert(\operator{1}-\operator{P}_{T\mathcal{M}_{\text{uMPS}}})(\operator{H}-H)\ket{\psi}\rVert$, which consists of a sum of quantities of comparable size which occasionally cancel each other. Large numerical errors can result from this cancellation.

\subsection{Details of the numerical integration schemes}
When the original Hamiltonian is invariant under time reversal, \textit{i.e.} $\operator{R}\operator{H}\operator{R}^{-1}=\operator{H}$, then the TDVP equations have been proven to respect this time-reversal symmetry in the previous section, provided that $\operator{R}\ket{\psi} \in \mathcal{M}$ for every $\ket{\psi}\in \mathcal{M}$. In quantum mechanics, the operator for time reversal is an antilinear operator. We  now assume that the local basis is chosen so that $\operator{R}$ can simply be implemented as complex conjugation $\operator{K}$, such that $\operator{K}\ket{\psi(A)}=\ket{\psi(\overline{A})}$. This ensures that the time-reversed state of any uMPS is still a uMPS. Time-reversal invariance of the Hamiltonian then requires that all entries of $\operator{h}$ are real. Note that, in the standard choice of basis for a spin $j$ system, $\operator{R}=\mathrm{e}^{-\ic \pi \operator{J}_{y}}\operator{K}$, so that mere complex conjugation is in fact a combination of time reversal and rotation of the spin state over an angle $\pi$ around the $y$-axis. However, we  ignore this subtlety and just refer to $\operator{K}$ as time reversal.

In the case of imaginary time evolution, we can restrict to real representations for $A$, $x^{\ast}$, and $B(x^{\ast})$ whenever the Hamiltonian is time reversal invariant so that $\operator{h}$ has only real entries. A simple implementation based on the Euler method was sketched in the main text. Even though this is only a first-order method, and it introduces large (second-order) errors, this is not an issue for an imaginary time evolution due to the inherent stability of the approach. For imaginary time evolution with the TDVP we expect a monotonically decreasing energy expectation value and as long as the step size $d\tau$ is small enough to reproduce this monotonic decrease, there is no need to change it. A decrease of time step should thus only be considered when higher-order effects result in an energy increase. The algorithm automatically slows down near the optimum since the norm $\lVert B^{i}(x^{\ast}) \ket{\partial_{i}\psi}\rVert$ approaches zero. The results for the $S=1$ Heisenberg antiferromagnet in the main text were obtained with a constant value $dt=0.1$. This is in sharp contrast with TEBD implementations where there is a Trotter error associated with the time step $dt$, and the step size should be decreased as we approach the optimum. This decrease of step size produces an additional slowing down that has a significant effect on the total number of iterations required to converge the state. Finally, after every Trotter step, the resulting state is a matrix product state with increased bond dimension and we have to truncate it. In infinite-size systems, the only possibility is to use a local truncation based on the Schmidt coefficients, which is not globally optimal. All of these points have been illustrated in the main text.

When using the TDVP to simulate real-time evolution, we are no longer  able to restrict to a real representation. The quantities $A$, $x^{\ast}$, and $B(x^{\ast})$  become complex, even when the operator $\operator{h}$ has only real entries. Nevertheless, the algorithm sketched in the main text remains valid in principle. However, the large (second order) errors that are introduced by the Euler method can now accumulate in time. A prime indicator of this is a drifting expectation value $\braket{\psi(t)|\operator{H}|\psi(t)}$ when the state $\ket{\psi}$ is evolved in time according to a time-independent Hamiltonian $\operator{H}$, whose expectation value should be conserved. The accumulation of systematic errors can be eliminated by implementing a numerical integrator for the Euler-Lagrange equations that respects the symplectic structure of the TDVP. However, the structure of the TDVP is much more complicated that the typical structure of classical dynamics with a separable Hamiltonian $H(q,p)=T(p)+V(q)$. In particular, the relation $H(\overline{A},A)$ is highly nonlinear and not separable. Consequently, none of the typical symplectic algorithms from classical dynamics can be applied to the TDVP.

If a set of differential equations is invariant under time reversal, it is a good policy to devise a numerical integration scheme that respects this time reversal symmetry. Numerical integration schemes that respect time reversal symmetry are called symmetric and share many nice properties with symplectic integration schemes such as a stable long-time behavior, a linear growth of the global error and a near-preservation of first integrals [E.~Hairer, C.~Lubich and G.~Wanner, ``Geometric Numerical Integration: Structure preserving algorithms for ordinary differential equations'', \emph{Springer Series in Computation Mathematics} {\bf 31}, Springer (2002).]. The following paragraphs  describes the details of a second order numerical integration scheme that respects time-reversal symmetry, although it can of course also be applied to Hamiltonians which are not invariant under time reversal, in which case it is a simple second-order numerical integrator. At any point, the calculation of the projection of $(\operator{H}-H)\ket{\psi}$ into the tangent plane $T_{A}\mathcal{M}$ in the point $\ket{\psi(A)}$ (\textit{i.e.}\ the determination of $x^{\ast}$) can be performed by the algorithm that was outlined in the main text.

\begin{figure}[h]
\includegraphics[width=8cm]{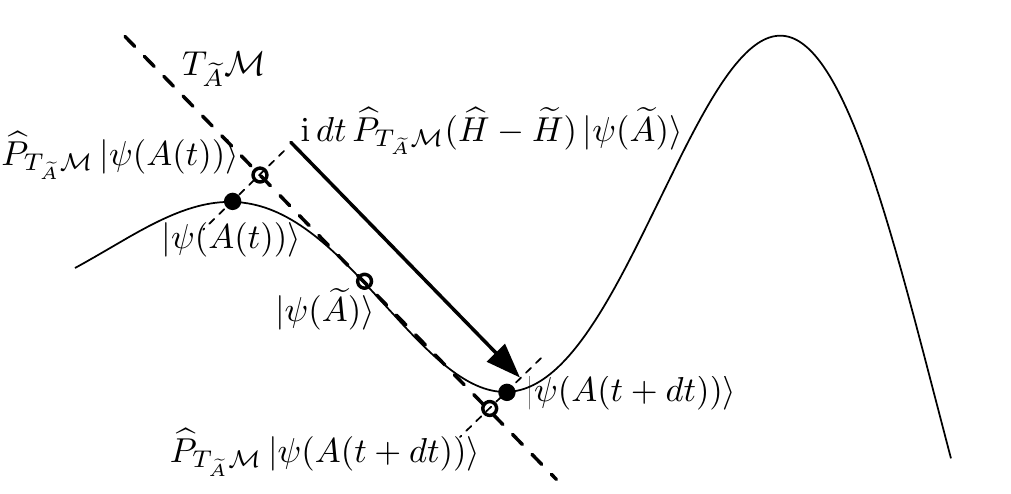}
\caption{Sketch of the location of the midpoint to be used in a symmetric integration scheme on a manifold}
\label{fig:midpoint}
\end{figure}

The main problem one encounters when trying to construct a symmetric integrator for differential equations on manifolds is that the tangent plane at $\ket{\psi(A(t))}$ and $\ket{\psi(A(t+dt))}$ are different. Most algorithms thus start with the determination of a midpoint $\ket{\psi(\widetilde{A})}$ such that the projection of $\operator{P}_{T_{\widetilde{A}}\mathcal{M}}\ket{\psi(A(t))}+\operator{P}_{T_{\widetilde{A}}\mathcal{M}}\ket{\psi(A(t+dt))}=0$, with $T_{\widetilde{A}}\mathcal{M}$ the tangent plane in the midpoint. This relation only specifies the location of the midpoint, but not the relation between $A(t)$ and $A(t+dt)$. In principle, the midpoint can be combined with any integration scheme, where we use the midpoint to calculate the variation $B(\widetilde{x}^{\ast})$. If we thus use the simple Euler step, we obtain the additional relation $\operator{P}_{T_{\widetilde{A}}\mathcal{M}}\ket{\psi(A(t+dt))}-\operator{P}_{T_{\widetilde{A}}\mathcal{M}}\ket{\psi(A(t))}=\ic\,dt\, \operator{P}_{T_{\widetilde{A}}\mathcal{M}}(\operator{H}-\widetilde{H})\ket{\psi(\widetilde{A})}$ where $\widetilde{H}=H(\overline{\widetilde{A}},\widetilde{A})$ is the energy in the midpoint. This is sketched in Figure~\ref{fig:midpoint}. This specific combination of midpoint with Euler step immediately tells us that 
$A(t+dt)=\widetilde{A}+\ic\, dt/2\, B(\widetilde{x}^{\ast})$, where $\widetilde{x}^{\ast}$ is the set of parameters such that $B^{i}(\widetilde{x}^{\ast})\ket{\partial_{i} \psi(\widetilde{A})} = \operator{P}_{T_{\widetilde{A}}\mathcal{M}} (\operator{H}-\widetilde{H})\ket{\psi(\widetilde{A})}$. Note that the representation of the variations $B$ itself depends on the current value of $A$, a subtlety which we  capture by simply introducing a suitable notation for the argument $x$. A similar reasoning leads to the conclusion that the midpoint is implicitly defined by $A(t)=\widetilde{A}-\ic\, dt/2\, B(\widetilde{x}^{\ast})$. However, we also have to take into account the normalization preservation and gauge fixing that is applied in every step. Instead of setting $A(t+dt)=\widetilde{A}+\ic\, dt/2\, B(\widetilde{x}^{\ast})$, we  find a scalar constant $c$ and a $D\times D$ matrix $G\in \mathsf{GL(D)}$ such that $A^{s}(t+dt)= c G [\widetilde{A}^{s}+\ic\, dt/2\, B^{s}(\widetilde{x}^{\ast})] G^{-1}$, where $c$ and $G$ are chosen such that $E^{A(t+dt)}_{A(t+dt)}$ has $1$ as largest eigenvalue, and satisfied a prescribed gauge fixing condition, such as the symmetric gauge defined in the previous subsection. Analogously, we also have to look for (different) $c$ and $G$ in the defining relation
$c^{-1} G^{-1} A^{s}(t) G=\widetilde{A}^{s}-\ic\, dt/2\, B^{s}(\widetilde{x}^{\ast})$, which are chosen such that $c^{-1} G^{-1} A^{s}(t) G-\widetilde{A}^{s}$ is compatible with the gauge fixing constraints that are built into the representation $B(\widetilde{x})$. Since every $B(\widetilde{x})$ satisfies $\rbra{\widetilde{l}} E^{B(\widetilde{x})}_{\widetilde{A}}=0$, we have to tune $c$ and $G$ such that 
\begin{displaymath}
\rbra{\widetilde{l}} E^{c^{-1} G^{-1} A(t) G-\widetilde{A}}_{\widetilde{A}}=0 \qquad \Rightarrow \rbra{\widetilde{l} G^{-1}} E^{A(t)}_{\widetilde{A}}=c \rbra{\widetilde{l}G^{-1}}.
\end{displaymath}
Put differently, $c$ is the largest eigenvalue of $E^{A(t)}_{\widetilde{A}}$ and $G$ is chosen such that $\rbra{\widetilde{l} G^{-1}}$ is the corresponding left eigenvector, where $\rbra{\widetilde{l}}$ is the eigenvector of $E^{\widetilde{A}}_{\widetilde{A}}$ corresponding to eigenvalue $1$. Since we cannot solve the resulting implicit relation $c^{-1} G^{-1} A^{s}(t) G=\widetilde{A}^{s}-\ic\, dt/2\, B^{s}(\widetilde{x}^{\ast})$ exactly, we  have to devise a numerical scheme to determine $\widetilde{A}$. The resulting algorithm doesn't satisfy time reversal symmetry exactly, but only up to the accuracy of the numerical determination of $\widetilde{A}$, which can be near machine precision.

We can try to solve the implicit relation for the midpoint by a simple error correct strategy. We choose as an initial guess $\widetilde{A}_{0} \sim A(t)+\ic\, dt/2 \, B(x^{\ast}(t))$, where the `similarity sign' is used to indicate that $\widetilde{A}_{0}$ has already been transformed in order to satisfy norm and gauge fixing constraints. Having a guess $\widetilde{A}_{n}$, we can iteratively try to improve it as follows. We calculate the difference $\widetilde{dA}^{s}_{n}=c_{n}^{-1} G_{n}^{-1}A^{s}(t) G_{n}-\widetilde{A}^{s}_{n}+\ic\,dt/2 B(\widetilde{x}^{\ast}_{n})$, where $c_{n}$ and $G_{n}$ are chosen such that $\rbra{\widetilde{l}_{n}G_{n}^{-1}} E^{A(t)}_{\widetilde{A}_{n}}=c_{n} \rbra{\widetilde{l}_{n} G^{-1}_{n}}$. We then set $\widetilde{A}_{n+1}\sim\widetilde{A}_{n}+\widetilde{dA}_{n}$ and repeat this process. At any point in the iteration, we can measure the size of the correction as $\lVert \widetilde{dA}_{n}^{i}\ket{\partial_{i}\psi(\widetilde{A}_{n})}\rVert =|\mathbb{Z}|^{1/2}\rbraket{\widetilde{l}_{n}| E^{\widetilde{dA}_{n}}_{\widetilde{dA}_{n}}|\widetilde{r}_{n}}^{1/2}$. As argued in the previous subsection, we can safely omit the overal $|\mathbb{Z}|^{1/2}$ in order to obtain a local measure $\zeta= \rbraket{\widetilde{l}_{n}|E^{\widetilde{dA}_{n}}_{\widetilde{dA}_{n}}|\widetilde{r}_{n}}^{1/2}$. When $\zeta$ dives below a tolerance level that can be chosen near-machine precision, we can stop the iteration. When the chosen time step is not too big --- for example $dt\approx 0.01$ --- this algorithm converges in a few (less then 20) iteration steps. Better strategies in terms of higher order iterative solvers for non-linear equations can be devised.

 The general outline of an algorithm for real time evolution is thus:
\begin{enumerate}
\item iteratively determine the midpoint from $A(t) \sim \widetilde{A} -\ic\, dt/2\, B(\widetilde{x}^{\ast})$
\item set $A(t+dt)\sim \widetilde{A} + \ic\, dt/2\, B(\widetilde{x}^{\ast})$
\end{enumerate}
Note that all operations can be implemented in $\mathcal{O}(D^{3})$ computation time. We  of course use iterative eigensolvers to determine the eigenvalues and eigenvectors of $E^{A(t)}_{\widetilde{A}}$, $E^{\widetilde{A}}_{\widetilde{A}}$ and $E^{A(t+dt)}_{A(t+dt)}$. 

The midpoints $\ket{\widetilde{A}}$ can in fact be interpreted as $\ket{A(t+dt/2)}$. Thus the algorithm produces twice the resolution as initially requested. However, it is only (approximately) time reversal invariant after an integral number of steps $dt$. Furthermore, since the (backwards) Euler method is used to step from $A(t)$ to $A(t+dt/2)$, the error in this step is expected to be $\mathcal{O}(dt^{2}/4)$. Similarly, the error in the step $A(t+dt/2)$ to $A(t+dt)$ is expected to be of the same order. Nevertheless, the resulting step from $A(t)$ to $A(t+dt)$ is correct up to second order, and the error is actually $\mathcal{O}(dt^{4})$ because odd-powered effects are forbidden by the symmetry of the construction. Higher order errors are obtainable by combining the midpoint construction with more advanced Runge-Kutta schemes. 

We can once again compare this implementation of the TDVP with a real time iTEBD simulation. Not only does the TDVP-based algorithm for real time evolution have the same advantages as the imaginary time algorithm --- conservation of translational invariance and internal symmetries --- the (approximate) time-reversal invariance makes the algorithm extremely stable over longer simulation times. In principle, the first step in a real time iTEBD implementation is even better, because evolving over a small time using a Trotter decomposition is a symplectic operation. However, since iTEBD takes the state outside the manifold of MPSs with fixed bond dimensions, this step is followed by a truncation that breaks both the symplectic symmetry and the time reversal symmetry. 

\begin{figure}[h]
\includegraphics[width=0.5\textwidth]{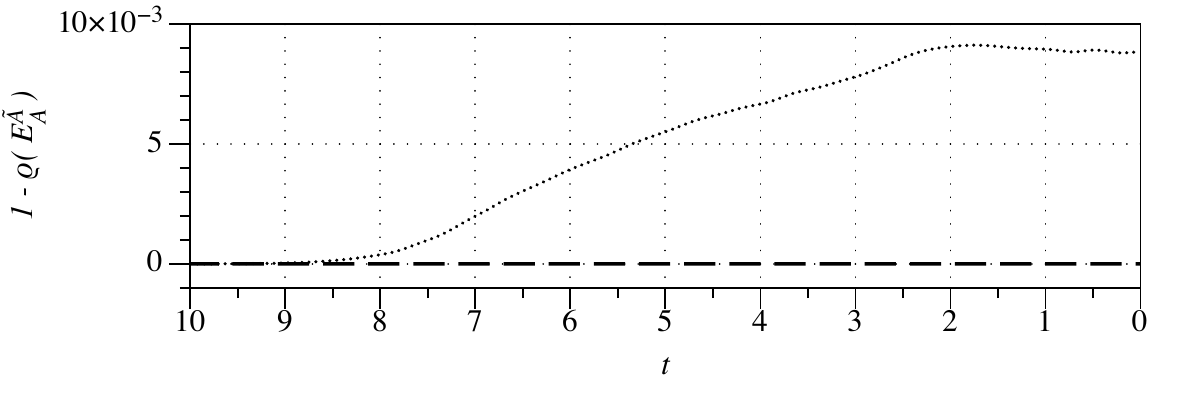}
\caption{Comparison of time reversal invariance in a TDVP (dashed lines) and TEBD (dotted lines) simulation. Illustrated is (one minus) the ground state fidelity $d(\overline{A}(t),\widetilde{A}(t))$, where $\widetilde{A}(t)=\overline{A'(t_{f}-t)}$ ($t_{f}=10$, $dt=5\times 10^{-3}$, $\zeta=10^{-10}$ (see text))}
\label{fig:timereversal}
\end{figure}

After having applied the real time evolution algorithm to evolve a state $\ket{\psi(A)}$ at $t=0$ to $\ket{\psi(A(t))}$ at time $t$ for a total time $t_{f}$, we can explicitly apply the time reversal operator $\operator{K}$ to the final state $\ket{\psi(A(t_{f}))}$ and use the resulting state $\ket{\psi(\overline{A(t_{f})})}$ as the initial value for a new time evolution over a total time $t_{f}$, resulting in states $\ket{\psi(A'(t)}$ at time $t$ with thus $A'(0)=\overline{A(t_{f})}$. We should then compare the states $\operator{K}\ket{\psi(A'(t_{f}-t))}=\ket{\psi(\widetilde{A}(t))}$ [with $\widetilde{A}(t)=\overline{A'(t_{f}-t)}$] to the states $\ket{\psi(A(t))}$. By using the results from the previous subsection, we know that the equality between two uMPS $\ket{\psi(A)}$ and $\ket{\psi(\widetilde{A})}$ can be measured by computing the ground state fidelity per site $d(\overline{A},\widetilde{A})=\rho(E^{\widetilde{A}}_{A})$, which is one for equivalent uMPS and will be smaller than one for states that differ. Fig.~\ref{fig:timereversal} compares the time-reversal symmetric integrator in this subsection with results a TEBD-based algorithm, for the matrices $A(t)$ were taken from the results in Fig.~2 in the main text.

\section{Dynamical expansion of the variational manifold}
Both for real and imaginary time evolution, we have introduced the local error measure $\epsilon=|\mathbb{Z}|^{-1/2} \lVert(\operator{1}-\operator{P}_{T\mathcal{M}_{\text{uMPS}}})(\operator{H}-H)\ket{\psi}\rVert$ that captures the tendency of the exact evolution to move away from the manifold $\mathcal{M}_{\text{uMPS}}$. If this quantity exceeds a given tolerance value, we might try to reduce the error by expanding the variational class. For uMPS with bond dimension $D$, we can expand $\mathcal{M}_{\text{uMPS}(D)}$ by increasing the bond dimension to some value $\widetilde{D}$. If the state $\ket{\psi}\in\mathcal{M}_{\text{uMPS}(D)}$ at some point is a uMPS with $D\times D$ matrices $A^{s}$, we can try to better approximate the exact evolution by embedding at the next iteration this state into a larger manifold $\mathcal{M}_{\text{uMPS}(\widetilde{D})}$ by defining new $\widetilde{D}\times \widetilde{D}$ matrices $\widetilde{A}^{s}$ with $\widetilde{D}\geq D$ and
\begin{equation}
\widetilde{A}^{s}=\begin{bmatrix} A^{s} &0\\ 0 &0\end{bmatrix}+dA^{s},
\end{equation}
where the variation $dA^{s}$ should be proportional to the chosen time step $dt$, and is given by
\begin{equation}
dA^{s}=\begin{bmatrix} dA^{s}_{00} & dA_{01}^{s}\\ dA_{10}^{s} & dA_{11}^{s}\end{bmatrix}.
\end{equation}
The first order effect of this evolution step is given by
\begin{equation}
\ket{d\psi_{0}}=\sum_{n\in\mathbb{Z}}\operator{T}^{n} \left(\sum_{\{s_{j}\}=1}^{d}v_{\mathrm{L}}^{\dagger}( \cdots A^{s_{-1}}(dA_{00})^{s_{0}}A^{s_{+1}}\cdots)v_{\mathrm{R}} \ket{\ldots s_{-1}s_{0}s_{+1}\ldots}\right).
\end{equation}
which is equivalent to the evolution in the original manifold. The newly introduced degrees of freedom do not seem to appear at first order, which is a consequence of the intrinsic nature of the TDVP not to leave the manifold in which it is defined. However, by choosing $dA_{01}$ and $dA_{10}$ proportional to $dt^{1/2}$, they do in fact generate a first order contribution
\begin{equation}
\ket{d\psi_{1}}=\sum_{n\in\mathbb{Z}}\operator{T}^{n} \left(\sum_{\{s_{j}\}=1}^{d}v_{\mathrm{L}}^{\dagger}( \cdots A^{s_{-2}}A^{s_{-1}}(dA_{01})^{s_{0}}(dA_{10})^{s_{+1}}A^{s_{+2}}\cdots)v_{\mathrm{R}} \ket{\ldots s_{-2}s_{-1}s_{0}s_{+1}s_{+2}\ldots}\right).\label{eq:secondorderpsi}
\end{equation}
Clearly the increased bond dimension allows for an action on two neighboring sites, which  increases the entanglement entropy from some initial value smaller than $\log(D)$ to some new value that is smaller than the new bound $\log( \widetilde{D})$. Note that $dA^{s}_{11}$ can only appear in first order if it is of order $\mathcal{O}(dt^{0})=\mathcal{O}(1)$. At this point, $dA^{s}_{11}$ can appear any number of times as a string in between $dA^{s}_{01}$ and $dA^{s}_{10}$. Thus $dA^{s}_{11}$ induces a non-trivial affect on any number of sites, with a minimum of three sites. It is thus not useful for nearest-neighbor Hamiltonians, and is henceforth set to zero: $dA_{11}=0$. We  show below that, by using the correct gauge conditions, the overlap between the exact evolution vector for a nearest neighbor Hamiltonian and such a vector with action on more than two sites is exactly zero. The freedom in $dA^{s}_{11}$ can possibly be useful in Hamiltonians with long range interactions. 

If we try to generalize the geometric strategy of the TDVP, we should look for optimal matrices $dA^{s}_{00}$, $dA^{s}_{01}$ and $dA^{s}_{10}$ such that 
\begin{equation}
\lVert dt (\operator{H}-H)\ket{\psi} - \ket{d\psi_{0}}-\ket{d\psi_{1}} \rVert
\end{equation}
is minimized. This is a complicated expression that couples the three sets of parameters $dA^{s}_{00}$, $dA^{s}_{01}$ and $dA^{s}_{10}$. Here too, we can apply infinitesimal gauge transformations
\begin{align*}
G&=\begin{bmatrix} 1+ \varepsilon X_{00} & \varepsilon ^{1/2} X_{01} \\ \varepsilon ^{1/2} X_{10} & 1+ \varepsilon X_{11}\end{bmatrix} 
\end{align*}
from which we obtain up to $\mathcal{O}(\varepsilon)$
\begin{align*}
 G \begin{bmatrix} A^{s} & 0 \\ 0 & 0\end{bmatrix}
G^{-1}= \begin{bmatrix} A^{s} & 0 \\ 0 & 0\end{bmatrix} + dA_{X}\quad \text{with}\quad dA_{X}= \begin{bmatrix} \varepsilon [X_{00},A^{s}]+\frac{\varepsilon}{2} \{ X_{01}X_{10}, A^{s}\}& -\varepsilon^{1/2} A^{s} X_{01} \\ \varepsilon^{1/2} X_{10} A^{s} & -\varepsilon X_{10} A^{s} X_{01}\end{bmatrix}.
\end{align*}
Any $dA$ is thus gauge equivalent to $dA+dA_{X}$. Note the absence of $X_{11}$ at order $\varepsilon $. It is also important to note that exploiting gauge freedom to reduce the number of independent parameters  mixes the off-diagonal blocks with the diagonal blocks. Consequently, we cannot generally treat the optimization with respect to $dA_{00}$ independently from the optimization with respect to $dA_{01}$ and $dA_{10}$. For evaluating $\braket{d\psi_{0}|d\psi_{0}}$ and $\braket{d\psi_{0}|(\operator{H}-H)|\psi}$, we can use the results from the main text. The corresponding relations involving $\ket{d\psi_{1}}$ are given by
\begin{multline}
\braket{d\psi_{1}|d\psi_{1}}=
|\mathbb{Z}|\Big[\rbraket{l|E^{dA_{1}}_{dA_{1}}|r}+\rbraket{l|E^{AdA_{1}}_{dA_{1}A}|r}+\rbraket{l|E^{dA_{1}A}_{AdA_{1}}|r}-3\rbraket{l|E^{AA}_{dA_{1}}|r}\rbraket{l|E^{dA_{1}}_{AA}|r}\\+\rbraket{l|E^{AA}_{dA_{1}}(1-E)^{-1}E^{dA_{1}}_{AA}|r}+ \rbraket{l|E^{dA_{1}}_{AA}(1-E)^{-1}E^{AA}_{dA_{1}}|r}
+|\mathbb{Z}|\rbraket{l|E^{AA}_{dA_{1}}|r}\rbraket{l|E^{dA_{1}}_{AA}|r}\Big],
\end{multline}
\begin{multline}
\braket{d\psi_{0}|d\psi_{1}}=
|\mathbb{Z}|\Big[\rbraket{l|E^{dA_{1}}_{AdA_{00}}|r}+\rbraket{l|E^{dA_{1}}_{dA_{00}A}|r}-2\rbraket{l|E^{A}_{dA_{00}}|r}\rbraket{l|E^{dA_{1}}_{AA}|r}\\+\rbraket{l|E^{A}_{dA_{00}}(1-E)^{-1}E^{dA_{1}}_{AA}|r}+ \rbraket{l|E^{dA_{1}}_{AA}(1-E)^{-1}E^{A}_{dA_{00}}|r}
+|\mathbb{Z}|\rbraket{l|E^{A}_{dA_{00}}|r}\rbraket{l|E^{dA_{1}}_{AA}|r}\Big]
\end{multline}
and
\begin{multline}
\braket{d\psi_{1}|\operator{H}-H|\psi}=|\mathbb{Z}|\Big[\rbraket{l|E^{C}_{dA_{1}}|r}+\rbraket{l|E^{AC}_{dA_{1}A}|r}+\rbraket{l|E^{CA}_{AdA_{1}}|r}\\+\rbraket{l|E^{AA}_{dA_{1}}(1-E)^{-1}E^{C}_{AA}|r}+ \rbraket{l|E^{C}_{AA}(1-E)^{-1}E^{AA}_{dA_{1}}|r}\Big],
\end{multline}
where $dA_{1}^{st}=dA_{01}^{s}dA_{10}^{t}$. While finding the simultaneous optimum for $dA_{00}$, $dA_{01}$ and $dA_{10}$ from these equations appears to be a very complicated task, we can exploit the gauge freedom to eliminate most terms. We can impose $\rbra{l} E^{dA_{01}}_{A}=0$ and $E^{dA_{10}}_{A}\rket{r}=0$ by exploiting the $D\times(\widetilde{D}-D)$ parameters in $X_{01}$ and the $(\tilde{D}-D)\times D$ parameters in $X_{10}$ respectively. This automatically removes all non-local terms from $\braket{d\psi_{1}|d\psi_{1}}$ and $\braket{d\psi_{1}|\operator{H}-H|\psi}$, \textit{i.e.}
\begin{align*}
\braket{d\psi_{1}|d\psi_{1}}&=|\mathbb{Z}|\rbraket{l|E^{dA_{1}}_{dA_{1}}|r},
&\braket{d\psi_{1}|\operator{H}-H|\psi}&=|\mathbb{Z}|\rbraket{l|E^{C}_{dA_{1}}|r}.
\end{align*}
In addition, this choice of gauge also renders $\braket{\psi|d\psi_{1}}=0$ and $\braket{d\psi_{0}|d\psi_{1}}=0$ for any $dA_{00}$. This choice of gauge thus ensures that $\ket{d\psi_{1}}$ is orthogonal to both the state $\ket{\psi}$ as well as to the complete tangent plane $T_{A}\mathcal{M}_{\text{uMPS}(D)}$ and only captures effects that act on two sites at once. These effects cannot be captured by $\ket{d\psi_{0}}\in T_{A}\mathcal{M}_{\text{uMPS}(D)}$, where only a single site is modified. The optimization problem for $dA_{00}$ completely decouples from the optimization problem for $dA_{01}$ and $dA_{10}$, and can be solved as described in the main text. Note that we still have the $D^{2}-1$ parameters in $X_{00}$ together with global norm conservation $\braket{\psi|d\psi_{0}}+\braket{\psi|d\psi_{1}}=\braket{\psi|d\psi_{0}}=0$ to fix the gauge of $dA_{00}$ as described in the main discussion. Furthermore, as stated above, with this choice of gauge for $dA_{01}$ and $dA_{10}$, the tangent vectors with nonzero choice of $dA_{11}$ are orthogonal to the evolution vector $(\operator{H}-H)\ket{\psi}$ for any nearest neighbour Hamiltonian.

For the optimization problem in $dA_{01}$ and $dA_{10}$, we can use a parameterization that is based on the definitions in the first subsection of this section. We define $B_{01}^{s}(x)=l^{-1/2}V_{L}^{s} x$ with $x$ a $(D-1)d\times (\widetilde{D}-D)$ matrix of independent components, and analogously $B_{10}^{s}(y)=y V_{R}^{s} r^{-1/2}$ with $y$ a $(\widetilde{D}-D)\times (D-1)d$ matrix of independent components. We then define $B_{1}^{st}(x,y)=B_{01}^{s}(x) B_{10}^{t}(y)$ in order to find
\begin{align*}
\rbraket{l|E^{B_{1}(x,y)}_{B_{1}(x,y)}|r}&=\mathrm{tr}\left[ (xy) (xy)^{\dagger} \right],&\rbraket{l|E^{C}_{B_{1}(x,y)}|r}&=\mathrm{tr}\left[G (xy)^{\dagger}\right],
\end{align*}
with $G=\sum_{s,t=1}^{d} (V^{s}_{L})^{\dagger} l^{1/2} C^{st} r^{1/2} (V^{t}_{R})^{\dagger}$. Since we have to minimize
\begin{align*}
\mathrm{tr}\left[(x y) (x y)^{\dagger} \right] - \mathrm{tr}\left[G (x y)^{\dagger} \right] - \mathrm{tr}\left[(x y)  G^{\dagger} \right],
\end{align*}
we are looking for the optimal matrix $xy$ of rank $\widetilde{D}-D$ such that $\lVert xy - G  \rVert$ is minimized, with $\lVert \cdot \rVert$ the Hilbert-Schmidt norm. The best approximation $(x,y)$can be found by performing a singular value decomposition of the $(d-1)D\times (d-1)D$ matrix $ G$ and retaining the largest $\widetilde{D}-D$ singular values. If $\widetilde{D}=d D$, we can always capture one step in the action of the Hamiltonian exactly, which is also the case with the TEBD algorithm.

This construction is very useful in combination with imaginary time evolution in order to correctly converge a random initial state to the optimal state within the variational manifold. Rather then starting from a random uMPS, it is better to use a previous optimal uMPS $A$ at a smaller bond dimension $D$ and combine it with the construction above to form an initial state $\widetilde{A}$ for the simulation at the new bond dimension $\widetilde{D}$.  Note that, if $A$ represents the optimal solution at $D$, we can choose $dA_{00}^{s}=0$, or thus $\ket{d\psi_{0}}=0$. This approach avoids a particular problem of the imaginary time evolution with the TDVP: For large $D$, the current TDVP implementation is susceptible to converging some random initial states to a state where $E^{A}_{A}$ has more then a single eigenvalue $1$. At this point, the formulas for the gradient and the gram matrix in the main text are no longer valid and incorrect application of these formulas often results in the algorithm getting trapped. While we could of course implement more advanced formula's that take this scenario into account and with which the implementation would be able to continue the convergence process, using the aforementioned construction avoids the occurrence of this problem completely. In addition, very few iterations are needed at the large values of $D$, since the error $\epsilon$ of the previous result at slightly smaller $D$ is already very small. Clearly, this is the preferred approach.

In principle, the same construction can be used in combination with real time evolution. After quantum quenches, the entanglement entropy  typically increases and so does the error measure $\epsilon$. When $\epsilon$ increases beyond a certain tolerance level, we can choose to use the above construction to dynamically increase $D$, which  brings $\epsilon$ back to an acceptable level. However, since this whole step is only well defined at first order, it cannot be included in a higher order numerical integrator. The dynamic expansion of the variational manifold should thus be used carefully in real time evolution, since it also breaks the time reversal symmetry.

Finally, we can also use the results from this section for another purpose. Since $\ket{d\psi_{0}}$ captures the projection of $(\operator{H}-H)\ket{\psi}$ in the tangent plane $T\mathcal{M}$, it is indeed equal to $B^{i}(x^{\ast})\ket{\partial_{i}\psi}=\operator{P}_{T_{A}\mathcal{M}_{\text{uMPS}(D)}}(\operator{H}-H)\ket{\psi}$. If we choose $\tilde{D}=d D$, then $\ket{d\psi_{1}}$ can exactly capture the remnant $(\operator{1}-\operator{P}_{T_{A}\mathcal{M}_{\text{uMPS}(D)}})(\operator{H}-H)\ket{\psi}$ that is orthogonal to the tangent plane $T_{A}\mathcal{M}_{\text{uMPS}(D)}$. We can use this to solve the following problems. If the exact evolution is almost exactly captured in the tangent plane $T_{A}\mathcal{M}_{\text{uMPS}(D)}$, the error measure $\epsilon$  has to be calculated as the square root of the difference of two `equally large' numbers $\Delta H^{2}$ and $\lVert B^{i}(x^{\ast})\ket{\partial_{i}\psi(A)}\rVert^{2}$. It is then numerically much more convenient to calculate $\epsilon$ as $\epsilon= |\mathbb{Z}|^{-1/2} \lVert \ket{d\psi_{2}} \rVert = \rbraket{l|E^{dA_{1}}_{dA_{1}}|r}^{1/2}$, which can completely be constructed without having to subtract large numbers with small differences. In addition, the expression for $\Delta H^{2}$ given in Eq.~\eqref{eq:dh} also contains $4$ contributions which might almost cancel each other if $\ket{\psi}$ is very close to an eigenstate of $\operator{H}$. This too can produce large numerical errors, and it is better to calculate $\Delta H$ as $\Delta H=\sqrt{\braket{\psi_{0}|\psi_{0}}+\braket{\psi_{1}|\psi_{1}}}$ where the square root contains two positive numbers. 

\section{Time-dependent variational principle for generic MPS}
We conclude this supplementary material by investigating whether new efficient algorithms can be obtained from applying the TDVP to generic (non-uniform and finite-size) MPS. The TDVP equation for finite size MPS was already written down in a different format in [J.J.~Dorando, J.~Hachmann, G.K.L.~Chan, {J. Chem. Phys.} {\bf 130}, 184111 (2009)], but the resulting equation was not further investigated. We define a MPS with open boundary conditions as
\begin{equation}
\ket{\psi[A(n)]}=\sum_{\{s_{n}\}=1}^{d} v_{\mathrm{L}}^{\dagger} A^{s_{1}}(1) A^{s_{2}}(2) \cdots A^{s_{L}}(L) v_{\mathrm{R}} \ket{s_{1}s_{2}\ldots s_{L}}
\end{equation}
with site-dependent matrices $A^{s}(n)$ having site-dependent dimensions $D_{n-1}\times D_{n}$. It is always possible to absorb $v_{\mathrm{L}}^{\dagger}$ into $A^{s_{1}}(1)$ and set $D_{0}=1$; similarly we can absorb $v_{\mathrm{R}}$ into $A^{s_{L}}(L)$ and set $D_{L}=1$. This completely eliminates the need for the boundary vectors, and they are henceforth  omitted. We can now define the series $l(n)$ and $r(n)$ (with $n=0,\ldots,L$) of site-dependent $D_{n}\times D_{n}$ density matrices for the auxiliary system through $l(0)=1=r(L)$ and
\begin{align}
l(n)&=\sum_{s=1}^{d} (A^{s}(n))^{\dagger} l(n-1) A^{s}(n),&r(n)=\sum_{s=1}^{d} A^{s}(n+1) r(n+1) (A^{s}(n+1))^{\dagger}.
\end{align}
The norm of the state is then given by $r(0)=l(L)=\rbraket{l(n-1)| E^{A(n)}_{A(n)}|r(n)}$, which we require to be one. 

If we want to apply the TDVP to the class $\mathcal{M}_{\text{MPS}}=\{ \ket{\psi[A(n)]}\}$ of MPS of length $L$ with fixed bond dimensions $\{D_{n},n=1,\ldots,L\}$, we have to vary with respect of all entries in every tensor $A(n)$. We now denote a general variation of $A$ as $B^{i}$ with a collective index $i=(\alpha,s,\beta,n)$, such that the tangent space $T_{A(n)} \mathcal{M}_{\text{MPS}}$ is spanned by the states
\begin{equation}
B^{i}\ket{\partial_{i} \psi}=\sum_{n=1}^{L} \sum_{\{s_{n}\}=1}^{d} A^{s_{1}}(1) \cdots B^{s_{n}}(n) \cdots A^{s_{L}}(L) \ket{s_{1}\ldots s_{n}\ldots s_{L}}.
\end{equation}
We can easily calculate the metric $G_{\overline{\imath}j}=\braket{\partial_{\overline{\imath}}\psi|\partial_{j}\psi}$, but it is a complicated expression coupling all variations, so it would seem like an impossible task to invert this $L D^{2}d \times L D^{2} d$ matrix. However, as for the case of uMPSs, we can exploit the gauge invariance to remove all non-local couplings, \textit{i.e.} all contributions containing variations at different sites. Combining the particular choice of gauge fixing conditions which has this effect with a well chosen linear parametrization yields an effective metric that is the unit matrix.

Firstly, we note that by choosing $B^{s}(n) = A^{s}(n)/L$, we have $B^{i} \ket{\partial_{i} \psi(A)}=\ket{\psi(A)}$. We  thus have to project onto norm-preserving variations. Secondly, we now have a site dependent gauge freedom $A^{s}(n)\leftarrow G(n-1) A^{s}(n) G(n)^{-1}$ with $G(n)\in \mathsf{GL}(D_{n})$. From the first order effect of an infinitesimal transformation $G(n)=1+\varepsilon X(n)$, which is $A^{s}(n)\leftarrow A^{s}(n)+\varepsilon B^{s}_{X}(n)$ with $B_{X}^{s}(n) = X(n-1) A^{s}(n) - A^{s}(n) X(n)$,  we conclude that $B^{i}_{X} \ket{\partial_{i} \psi(A)}=0$ as can easily be checked explicitly. Thus any variation $B^{i}$ is gauge equivalent to $B^{i}+B^{i}_{X}$ for all $\{X(n)\in \mathbb{C}^{D_{n}\times D_{n}}\}$. 

We can fix the gauge of the variations (and ensure norm conservation) by requiring that either $\rbra{l(n-1)}E^{B(n)}_{A(n)}=0$ or $E^{B(n)}_{A(n)}\rket{r(n)}=0$ for every $n$, which we call left and right gauge fixing conditions. Both choices  reduce the metric to
\begin{equation}
\overline{B'}^{\overline{\imath}} G_{\overline{\imath},j} B^{j}= \sum_{n=1}^{L} \rbraket{l(n-1) | E^{B(n)}_{B'(n)}|r(n)}.
\end{equation}
If $\operator{H}$ is a nearest neighbor Hamiltonian $\operator{H}=\sum_{n=1}^{L-1}\operator{h}(n)$ with $\operator{h}(n)$ acting non-trivially only on sites $n$ and $n+1$, then the corresponding right hand side of the Euler-Lagrange equations reduces for the case of left gauge fixing to
\begin{equation}
\begin{split}
\overline{B}^{\overline{\imath}} \braket{\partial_{\overline{\imath}}|\operator{H}-H|\psi}=\sum_{n=1}^{L}\bigg[&\theta(n<L)\rbraket{l(n-1)|E^{C(n)}_{B(n) A(n+1)}|r(n+1)}\\
&+\theta(n>1)\rbraket{l(n-2)|E^{C(n-1)}_{A(n-1) B(n)}|r(n)}\qquad\\
&+ \theta(n>2)\sum_{m=1}^{n-2} \rbraket{l(m-1)|E^{C(m)}_{A(m)A(m+1)}\left(\prod_{k=m+2}^{n-1}E^{A(k)}_{A(k)}\right) E^{A(n)}_{B(n)} |r(n)}\bigg]
\end{split}\label{eq:gmpsgrad}
\end{equation}
where $\theta$ is a discrete Heaviside function that yields one if its argument is true and zero otherwise, and where $C^{st}(n)=\sum_{u,v=1}^{d} \braket{s,t|\operator{h}(n)-h(n)|u,v} A^{u}(n) A^{v}(n+1)$ and 
\begin{equation}
h(n)=\sum_{u,v,s,t=1}^{d}\braket{s,t|\operator{h}(n)|u,v} \mathrm{tr}\left[l(n-1) A^{u}(n) A^{v}(n+1) r(n+1) (A^{s}(n) A^{t}(n+1))^{\dagger}\right].
\end{equation}
In case of right gauge fixing, the last term in Eq.~\eqref{eq:gmpsgrad} would be replaced by a term containing all contributions of the Hamiltonian acting to the right of $B(n)$. These terms are familiar from the variational sweeping algorithm for MPS, and it is well known how to construct them efficiently and iteratively. Clearly now, all variations $B(n)$ decouple from variations $B(n')$ at different sites $n'\neq n$. 

Analogously to the case of uMPS, it is easy to find a parameterization of $B[x]$ depending on $L$ matrices $x(n)$ ($n=1,\ldots,L)$ of size $(d D_{n-1} - D_{n})\times D_{n}$, such that $B[x](n)$ depends only on $x(n)$ and such that the effective Gram matrix becomes the unit matrix. We first define the $D_{n}\times d D_{n-1}$ matrix
\begin{equation}
[L(n)]_{\alpha,(s\beta)}= [ A^{s}(n)^{\dagger}l(n-1)^{1/2}]_{\alpha,\beta}
\end{equation}
and then construct a $d D_{n-1}\times (d D_{n-1}-D_{n})$ matrix $V_{L}(n)$ that contains an orthonormal basis for the null space of $L(n)$, \textit{i.e.} $L(n) V_{L}(n)=0$ and $V_{L}(n)^{\dagger}V_{L}(n)=1$, for all $n=1,\ldots,L$. We then use the representation
\begin{equation}
B^{s}[x](n)= l(n-1)^{-1/2} V_{L}^{s} x(n) r(n)^{-1/2}
\end{equation}
in order to obtain
\begin{equation}
\overline{B}^{\overline{\imath}}[x] G_{\overline{\imath}j} B^{j}[y]=\sum_{n=1}^{L} \mathrm{tr}\left[ x(n)^{\dagger} y(n)\right].
\end{equation}
Let us now outline all the necessary steps in a single iteration of a simple Euler implementation for imaginary time evolution:
\begin{enumerate}
\item Define $K(0)=0$ (scalar) and $K(1)=0$ ($D_{1}\times D_{1}$ matrix) and compute the $D_{n}\times D_{n}$ matrix $K(n)$ as $$K(n)=\sum_{s,t=1}^{d} A^{t}(n)^{\dagger}A^{s}(n-1)^{\dagger}l(n-2) C^{st}(n-1)+\sum_{s=1}^{d} A^{s}(n)^{\dagger}K(n-1) A^{s}(n)$$
for $n=2,\ldots,L-1$.
\item Define the $(d D_{n-1}-D_{n})\times D_{n}$ matrices $F(n)$ as
\begin{equation}
\begin{split}
F(n)=&\theta(n<L)\sum_{s,t=1}^{d}V^{s}_{L}(n)^{\dagger} l(n-1)^{1/2} C^{st}(n) r(n+1) A^{t}(n+1)^{\dagger} r(n)^{-1/2}\\
&+\theta(n>1) \sum_{s,t=1}^{d}V^{s}_{L}(n)^{\dagger}l(n-1)^{-1/2} A^{t}(n-1)^{\dagger} l(n-2) C^{ts}(n-1) r(n)^{1/2}\\
&+\theta(n>2) \sum_{s=1}^{d} V^{s}_{L}(n)^{\dagger} l(n-1)^{-1/2}K(n-1) A^{s}(n) r(n)^{1/2}.
\end{split}
\end{equation}
\item Set $x^{\ast}(n)=F(n)$.
\item Take a step $A^{s}(n,t+dt)= A^{s}(n)-dt B^{s}[x^{\ast}](n)$.
\item Set $l(0)=1$ and compute $l(n)=\sum_{s=1}^{d} (A^{s}(n))^{\dagger} l(n-1) A^{s}(n)$ for $n=1,\ldots,L$. Renormalize if necessary. Set $r(L)=1$ and compute $r(n)=\sum_{s=1}^{d} A^{s}(n+1) r(n+1) (A^{s}(n+1))^{\dagger}$ for $n=L-1,\ldots,0$. Also compute $h(n)$ and change gauge fixing if necessary. 
\item Compute $H=\sum_{n=1}^{L} h(n)$ and evaluate the step. Change $dt$ if necessary.
\end{enumerate}
Every step can be computed with computation complexity $\mathcal{O}(L D^{3})$, just as in the standard sweeping algorithm. However, unlike in the sweeping algorithm, the determination of the change $A^{s}(n)\leftarrow A^{s}(n)-dt B^{s}[x^{\ast}](n)$ --- even though only valid for $dt$ not too large --- is globally optimal and thus includes all correlations from the changes at every other site in the lattice. In addition, the variation $B^{s}[x^{\ast}](n)$ does not require to (iteratively) solve an eigenvalue equation but is determined directly from matrix multiplications of $\mathcal{O}(D^{3})$. Since the iterative solution of the the eigenvalue problem is the dominating contribution to the computation time for large $D$ in the traditional sweeping algorithm, it is interesting to see whether this implementation can outperform the traditional sweeping algorithm in convergence speed. Clearly, all the other aspects of this supplementary material (a time-reversal invariant integrator for real time evolution, dynamical expansion of the variational manifold, error measures, \ldots) can also be applied to the non-uniform, finite-size lattice. Note that, for lattices with periodic boundary conditions, the gauge freedom cannot be exploited to fully decouple variations at different sites, resulting in a much more complicated algorithm, which is also the case with other DMRG-related approaches for lattices with periodic boundary conditions.